\documentstyle[11pt]{article}
\newcommand{\blankline}{\vskip .3cm}
\newcommand{\f}{\begin{equation}}
\newcommand{\ff}{\end{equation}}

\begin{document}
%\rightline{\Large CGPG-95/5-2, IASSNS-95/30}
\vskip .3cm
\centerline{\LARGE Cosmology as a problem in critical phenomena}
\rm
\vskip.3cm
\centerline{Lee Smolin${}^\dagger$}
\vskip .3cm

 \centerline{\it   Center for Gravitational Physics and
Geometry, Department of Physics${}^*$}
\centerline{\it Pennsylvania State University,
University Park, Pa16802-6360, USA}
\centerline{and}
\centerline{\it   School of Natural Sciences, Institute for
Advanced Study}
\centerline{\it Princeton, New Jersey, 08540, USA}

 \vfill
%\centerline{May 7, 1995}
\vfill
\blankline
\centerline{ABSTRACT}
\blankline
\noindent
\blankline
\noindent
A measure of complexity which is suggested by these applications,
but which may also have application to other problems, is described.

\noindent
 \vfill
$\ \ {}^\dagger$  smolin@phys.psu.edu \ \ \  * permanent address
\eject

\section{Introduction}

Until recently, most attempts to construct theories
of physics and cosmology have begun with the point of view that
the universe is, in its fundamentals, not very complicated.
Unfortunately,
it seems that the world often frustrates our desire to understand
it simply: $\Omega$\footnote{The average density of
the universe in units of the amount needed for the characteristic
time scale of the cosmic expansion to be infinite.}
must have been very finely
tuned originally
to be close to one now, but the best evidence is that it is
now in fact measurably
less than one\cite{Omega-WNE,ColesEllis}.
The neutrinos are very light compared to every
other mass scale, but there is  evidence that
they are not exactly massless\cite{numass}, while the
proton and neutron have almost, but not exactly,
the same mass.  Similarly, the
cosmic microwave radiation that gives us
a snapshot of conditions when the universe was a thousand
times smaller is a black body to incredible precision, and is isotropic
to a precision of around a part in $10^5$ \cite{COBE}; while
at the present time,
surveys of the actual distribution of matter show a world which
has structure up to the largest scales that have been accurately
mapped\cite{structure}.

The last thirty years have indeed been
an incredibly surprising and
exciting time in cosmology and theoretical physics.  At the risk
of oversimplifying, it seems that our
attempts to model the universe on both cosmological
and microscopic scales are leading to the conclusion that the
universe is much more intricately structured than was imagined
in the nineteen-sixties.  There is always risk in generalizations,
but if one looks for them, it might be said that
three themes have emerged during
this period that characterize the direction in which we seem
to be headed in both cosmology and elementary particle theory.

{\bf Complexity}  On many different scales, we are discovering
that the universe is much more complex than we might have
expected based on earlier theoretical ideas.  At the largest scales, the
distribution of galaxies in space shows structure that was largely
unexpected\cite{structure}, whose origins are still
not satisfactorily understood.
Finally, as I will
describe, the galaxies themselves seem to be much more complex
than might have been expected.

At the smallest scales, with the discovery of the charmed,
bottom and top
quarks, and the tau leptons,
the number of fundamental particles has just about doubled since
1970.  But
we still have no understanding of the spectrum of fundamental
fermions, nor do we have
a theory that explains the eighteen or so parameters
in the standard model.  The spectra of masses and
mixing angles shows a complexity
that is rather puzzling, with up and down quarks quite light on the
hadron scale, while the others are spread over a range of masses up
to  almost $200 Gev$.  The pattern of mixing angles is
also rather complex, and we have to understand funny things like
why parity seems to be so well respected by all of
the interactions but one, which breaks it maximally, or
why $CP$ is broken, but
 just a bit.  Whatever pattern of symmetry
breaking is behind all of this, it is unlikely
to be simple.  The models of grand unification that are now being
considered are correspondingly
rather more complicated than the original
$SU(5)$ theory, that had to be discarded because proton decay,
if it takes place, is rarer than that theory naturally
predicted.  Unification
has turned out to be a harder problem than perhaps it
seemed in 1975, partly because the properties of the elementary
particles and forces are themselves so diverse.

Furthermore, it seems that the world on every scale larger than
the nucleus is much more complex, given the actual values of
the masses, coupling constants and mixing angles, than would be
the case were they to take most other
values\cite{Hoyle,CarrRees,BarrowTipler,Carter}.  For
example, the fact that there
are many different stable bound states of protons and neutrons
seems due to several coincidences in the values of these parameters.
It may even be said that the complexity of the world on astronomical
scales is to some extent a consequence of the complexity of the
spectrum of elementary particles and forces.  For most
other values
of these parameters the chemistry, atomic and nuclear physics and
astronomy of the world would be much simpler.

{\bf Hierarchies and approximate scale invariance}
In both cosmology and
elementary particle physics, the basic units of structure are
spread over many orders of magnitude in scale, and notions
of approximate scale invariance play an important role.
Perhaps the most basic unsolved problem in elementary
particle physics is the hierarchy problem, which is to explain
why there are such large ratios among the fundamental scales
in physics.  In the fundamental Planck units, the mass of the
proton is $10^{-19}$, the electron is three orders of magnitude
smaller, and the cosmological constant is at most $10^{-60}$.

Furthermore, the fact that the astronomical world shows structure
on such a wide range of scales is a direct consequence of this
hierarchy in fundamental physics.  The typical mass of a star
is given by
\f
M_{Chandra} = m_{proton} \left ( { m_{Planck} \over m_{proton}}
\right  )^{3},
\ff
while its lifetime is given by
\f
t_{star} \approx  \alpha \epsilon { m_{proton} \over m_{electron} }
\left ( {M_{Chandra} \over M } \right )^2  \
\left ( { m_{Planck} \over m_{proton}} \right  )^{3} t_{Planck}
\approx
\left ( {M_{Chandra} \over M } \right )^2  \
10^{10} \ years
\ff
where $\epsilon \approx .007$ is the fusion efficiency.

It has also been
estimated that the typical mass of a galaxy
must be\cite{CarrRees,BarrowTipler},
\f
M_{galactic}= m_{proton} \alpha^5
\sqrt{m_{proton} \over m_{electron}}
\left ( { m_{Planck} \over m_{proton}} \right  )^{3}
\ff

Thus, the hierarchical structure we see in astronomy,
with stars organized into much larger galaxies, which seem in
turn to be collected in still larger structures, is
actually a consequence
of the hierarchy among the scales in fundamental physics.

There is also evidence
 that approximate scale invariance
characterizes the distribution of galaxies in space, at least
over a certain range of scales\cite{peebles-book}.
In addition, the most
successful hypotheses for the initial fluctuations in mass
density that ultimately lead to the formation of the galaxies
and the large scale structure is that their distribution is
scale invariant\cite{peebles-book}.

Going back to the small scale structure, because the fundamental
length, the Planck length, is so small compared to the scales
of strong interaction physics, the ground state of elementary
particle physics is characterized by an approximate scale
invariance at all scales larger than $l_{Planck}$.  This has
led to important conceptual
tools in elementary particle theory,
such as the renormalization group and the analogy between
a quantum field theory and a statistical mechanical system
at a second order critical point.

{\bf Evolution}  The most important way in which twentieth
century cosmology differs from the Newtonian and
Aristotelian cosmologies is
that it is based on the understanding that the universe has
evolved dramatically over time.  Whatever happens concerning
the details of the very early universe and the problems of
structure formation, the successes of the big bang
model, together with the failure of the steady state theory, leave
us with a universe whose present state must be understood to
be the result of physical processes which occurred at earlier times,
when it was very much different.  Thus, cosmology has become
an historical science, in which a detailed story
of what happened at earlier
times has replaced the philosophical and {\it a priori}
speculations that
characterized most previous attempts at cosmology.

The notion of evolution has not so far played a correspondingly
central role in elementary particle physics.  This may, on
reflection, seem unnatural, given the close relationship
that is developing between particle physics and cosmology.
Certainly, one must wonder what the traditional notion that the
laws of physics represent timeless truths means in a universe
whose origin we can literally almost see.

In the body of these notes I will elaborate on
some implications of these three themes.  Before beginning,
however, some general comments are in order.

\subsection{Why critical phenomena may be important for
particle physics and cosmology}

Since the 1970's there has been a mutually fruitful interaction
between statistical mechanics and elementary particle physics,
based largely on the formal analogy between second order
phase transitions and the problem of renormalization in
quantum field theory.  At the root of this, however, is a deep
problem for elementary particle physics, for this analogy is
based on the fact that there is a fine tuning problem in quantum
field theory.  The parameters that specify the dynamics
must be precisely tuned as a function of the cutoff scale, if there are
to be interacting particles on scales much larger than the cutoff.
It has helped a great deal to understand that this problem is
analogous to the problem of tuning a statistical system to a critical
point to describe a second order phase transition, but it does not
solve the basic problem of why a fine tuning is needed in quantum
field theory.

We may note that as long as renormalization is thought of as
a mathematical process in which the cutoff energy scale is
taken to infinity, then the fine tuning problem I have been
speaking about is formal, as it concerns a technique
used to construct the theory, and does
not describe any phenomena in nature.
But one thing all at least partially
successful approaches to quantum gravity
agree about is that the Planck scale does
function as an effective short distance cutoff\cite{garay}.
For apparently
different reasons this is the case both in string theory and in the
nonperturbative approaches to diffeomorphism invariant
quantum field theories\footnote{Of course, many people have
hypothesized the existence of such a fundamental scale, what
is significant is that it comes out of these two approaches to
combining quantum theory and relativity without being
put in by hand.}.
Once there is a physical cutoff the analogy between statistical
mechanics and Euclidean quantum field theory becomes perfect
and the fine tuning problem becomes a physical problem.
It then becomes a problem of
physics, and of critical phenomena in particular, to understand why
our world has light particles in it.

More recently, a second domain of critical phenomena has come to
light in statistical physics, in which no fine
tuning is necessary\cite{BTW,per-soc}.  These are self-organized
critical systems, which are non-equilibrium systems that
spontaneously organized themselves in configurations characterized
by approximate scale invariance over a wide range of scales, without
the need for any precise tuning of parameters.  It is then natural
to ask whether such mechanisms, or some general
mechanism of self-organization,
might also play a role in elementary particle  physics, to explain
the fine tunings, and the existence of large hierarchies, that we now
must impose.

Critical phenomena associated with phase transitions have also
played a role in early universe cosmology.  The two best studied
ideas to explain the ultimate origin of the large scale structure,
inflation and cosmic strings, involve phase transitions
as the universe expands and cools.  Both of these
can lead to scale invariant distributions of initial fluctuations
of the type that seem necessary to explain the current data
about the large scale structure.  However, in spite of these
successes, there are indications that the models which have
been studied so far may not in the end account completely
for the large
scale structure that is seen.  The most important reason for this
is that, as I
mentioned in my opening, the evidence is more and more pointing
to an $\Omega$ less than one.
It is then natural to ask whether
the more recently studied self-organized critical phenomena
might play some role in the early history of the universe, and
whether this might provide an alternative framework for
understanding structure formation, and the origin of
approximate scale
invariance, in the large scale structure of the universe.

There have already been several proposals about how the statistical
mechanics of self-organized systems may play a role in
astrophysics.    There are conjectures that the spectra of
radiation coming from accretion disks around neutron stars
or black holes
might arise from self-organized critical
systems\cite{soc-accretion}.  In addition,
there are suggestions that spiral galaxies
may be described as stable non-equilibrium systems which are
self-organized by the action of certain feedback processes involving
star formation.  These examples suggest that there may be fruitful
scope for applying the physics of non-equilibrium and self-organized
systems to problems in astrophysics and cosmology.

But perhaps more generally, I would like to propose that there
must be a role for the physics of self-organized systems in
cosmology and particle physics, simply because of the fact that
it is highly non-trivial that the
universe is as organized as it is. If it is the case that for most
values of the parameters of particle physics and cosmology, and
most choices of initial conditions, the universe would be much
less varied and organized than it is presently, then there must
be some reason for this.  Given the incomplete
success of other hypotheses,
it perhaps is not inappropriate to begin to look for new ideas
about the choices of parameters and initial conditions
according to which the fact that the world is so organized
may turn out to be essential rather than accidental.

But, if we seek a scientific explanation for this circumstance,
then we have no recourse except in the physics of self-organized
systems.  The anthropic principle won't help us, for it assumes what
we want to explain, which is that the universe is
sufficiently intricately organized
and out of equilibrium that life may exist.  There is nothing outside
the universe, by definition, so any processes that have acted in the
past to organize it must be processes of self-organization.

Furthermore, due to the advances in the theory of self-organized
systems due to Per Bak and his collaborators, we now know that
self-organized systems are often critical systems, with
structure spread out in space and time over every available scale.
The fact that the distribution of matter in our universe is
approximately scale invariant over many orders of magnitude
suggests
that it may be fruitful to seek to apply some of the ideas and
strategies developed in the study of self-organized systems to
unsolved problems in cosmology and astrophysics.

These notes are then meant as an introduction to several
different problems in astrophysics and cosmology in which
critical phenomena might plausibly play an important role.
I begin in section
2 with the problem of the organization of spiral galaxies
and then in the next two sections describe the open
problems in our understanding of the large scale organization
of the observed universe.  All the facts presented in these
three sections will be familiar to astronomers, even if
the point of view may be
nonstandard\footnote{I apologize that, as these notes are intended
as an introduction to these areas, and not as a comprehensive
survey, no attempt has been made to provide a complete
set of references.}.
The last three then sections concern
ways in which critical phenomena or mechanisms of
self-organization may play a role in elementary particle
theory, quantum gravity and general relativity.

\section{Spiral galaxies as self-organized systems}

A good place to start the discussion is with the disks of
spiral galaxies, as this is one astrophysical
domain in which it is clear that non-equilibrium processes are
responsible for the formation and maintenance of structure.
For this reason, it is also the one domain of astronomy
that has been attacked in a serious way by physicists using
the tools of modern statistical physics such as percolation theory
and cellular automata.   In a series of very interesting papers,
Seiden, Schulman and Gerola constructed a theory of spiral
structure based on an understanding of star formation as
a certain kind of percolation process that spreads through the
disk of the galaxy\cite{IBMguys,ssg-morphology}.
To introduce the basic
ideas of their theory
I need first to review some of the basic facts about stars and
galaxies.

A spiral galaxy, such as our own, consists of a number of components
which are characterized by a surprising variety of structures
and processes.  The galaxy is
surrounded by a spherical halo consisting primarily of old stars,
as well as some unknown form of non-luminous matter.  This dark
matter seems to provide about $80-90 \% $ of the mass, but does not
otherwise participate in the energetics of the
galaxy.  For the moment we may leave to one side the
very interesting question of its
composition and origin.

Embedded in this halo is a disk consisting of gas, dust and stars
of all ages.  It is here that the dynamical processes that distinguish
a spiral galaxy take place, and this will be the primary focus of this
discussion.  In the center of the disk is a bulge, which, like the
halo, consists primarily of old stars.  In the galaxies we will be
concerned with here, the disk is much larger than the bulge.

The disk of a spiral galaxy seems to be a system which exists
in a steady state, far from equilibrium, which is maintained by
processes which cycle matter and energy among its various
components
\cite{spiralstructure,WyseSilk,clouds,elmegreen-triggered}
\cite{elmegreen-review,parravano}.
The evidence is that the rates of these flows are approximately
constant, averaged over the whole disk of the
galaxy.  Not
surprisingly, some astronomers have
proposed that there are
feedback processes that govern the rates of flows of these
cycles\cite{spiralstructure,parravano}.
To understand them we first must be familiar with
the basic components and processes that make it up.

Stars come in a range of masses, from about $1/10$ to
$100 M_{solar}$, where $M_{solar}= 2 \times 10^{33} grams$ is the
mass of our sun.  It is important to know that the luminosity
of a star increases like the cube of its mass, so that
the more massive stars dominate the energetics of the galactic
disks.  However, the lifetime decreases drastically, scaling
like $mass^{-2}$.  Because of these two facts, the stars of
different masses play very different roles in the system of
a galaxy.  One basic fact is that the brighter and more massive
stars radiate
predominantly in the ultraviolet, so that they appear blue, while
the less massive ones radiate primarily in the red.

Our understanding of the processes by which stars
 are formed is growing very rapidly at the
present time\cite{stars}.  What is
clear is that, at least in spiral galaxies like our own, stars form in
certain clouds of gas and dust called giant molecular clouds.  We
will speak about these shortly.  A second very important fact
is that the stars are created with a distribution of masses
which is approximately a power law.   This distribution is
called {\it the initial mass function}, or
IMF\cite{Salpeter,Scalo}.
Many more  lower mass stars are formed
originally;  there is an empirical power law, due to Salpeter, that
the number of stars born with mass between $m$ and $m+dm$
scales like $m^{-\gamma }$ with $\gamma$ a power
between $2$ and $3$.  There is  evidence for a cut off
on the low end, so that stars smaller than about $1/20$ of a solar
mass are rare.  There is also controversial evidence that the
powers are different for low mass and high mass stars, which would
suggest that they are formed in different
processes\cite{Larson}.

Astronomers have looked for evidence that this initial mass function
has varied over the lifetime of the galaxy or differs among galaxies;
none has so far been found\cite{Scalo}.

As a result of this, together with the fact that the more massive stars
live for short times, the population of stars is dominated by the
low mass stars.  But, where they are found,
the energetics is dominated by the
massive stars.

The lower mass stars have lifetimes comparable to the present
life of the universe ($10^{10}$ years).  When they burn out they
end up quietly as a white dwarf.  However, those stars more
massive than about $8 M_{solar}$ end as supernovae, by which
they expel all but about $1-3$ $M_{solar}$'s of their mass.
The supernovaE also contribute a great deal of energy to the
galaxy.    These massive stars live for much shorter periods, with
the time between formation and supernova typically on the
order of $10^7$ years.  As this is much less than the rotation
time of the galaxy (which is of order $10^8$ years), this means
that massive stars are found only in or
near regions where star formation
is taking place.

The spiral patterns one sees looking
at a galactic disk are primarily caused by the very bright,
massive stars.  As such, these patterns trace the process of
star formation.  The  disks apparently manifest spiral structure for
the life of a galaxy, which is at least $10^{10}$ years.
This is, at least in some galaxies, connected to the fact
that the star formation rate is
constant\footnote{The spirals with constant star formation
rate are type Sc, which have the largest ratio of the
size of the disk relative to the bulge.  In galaxies
with much larger bulges the total star formation rate
is now less than it was in the past.  This suggests
that a constant star formation rate is a property associated with the
disk.}.

There are other processes besides supernova by which stars
return matter to the interstellar medium of gas and dust out
of which they are born.  Massive stars evaporate a significant
portion of their mass, this is the primary origin of the dust.

The dust and the gas together make up  a clumpy medium
which collects at the midplane of the disk.  As a layer
of gas, there is growing evidence that the disk extends far
beyond the disk of stars.  In the inner region containing the
stars, the interstellar medium exists in several distinct
phases, with greatly varying
temperatures and pressures.  To understand the role of
the medium we need to describe these different
phases\cite{clouds,elmegreen-review}.

Most of the volume of the medium is taking up by a
very rarified
phase
of hot ionized gas, with temperatures of greater than
$10^5  K$.  These are regions that have been evacuated
and ionized
by the passage of shock waves from supernovae.  Next, going
down in temperature, is a phase
of warm gas, with temperatures on the order of $10^3$
degrees $K$ and densities on the order of one atom per $cm^3$.
The gas is primarily atomic hydrogen.

Embedded in this warm gas are denser clouds, which apparently
are continually condensing out of it.  These clouds range
from $10-100$ degree kelvin, with densities that range
inversely from one up to $10^4$ atoms per $cm^3$.  In
the denser and colder clouds there is a lot of dust, which
apparently plays a role shielding the cloud from the ultraviolet
light that would heat it.  Because of this shielding, the gas in
the clouds is molecular.  Not only is the hydrogen
in molecular form,
but an array of organic elements are found there, including
not only $CO$ and $NH_3$ but alcohols and larger organic
molecules.  Because of this these are called the giant
molecular clouds.  They have masses on the order of
$10^6$ solar masses, and diameters of a few light years.

The distribution of matter within these giant molecular
clouds is very irregular.  There are suggestions that they
have a filamentary structure; there are also suggestions
that the distribution of densities in them is scale invariant
up to large scales\cite{scalo-fractal}.

The giant molecular clouds play a key role in the galaxy because
it is in them that the stars form.

The most important thing to understand about the star formation
process in the giant molecular clouds is that it is
rather inefficient\cite{stars}.
This seems to be true for three reasons.  First, a star begins to
form when a cold and dense core of a cloud collapses.  At some
point its center is dense enough to ignite.  This fuels a wind, which
blows out from the star, or from an accretion disk surrounding it,
which blows away the matter around the star, cutting off the
accretion of matter onto the star.  It is likely that this feedback
process is responsible for the fact that the typical mass of a star
is in fact just right for nuclear burning.

Second, when massive stars,
form in a cloud, they heat it which after sufficient energy
has been deposited in the cloud, apparently curtails  further star
formation\footnote{Evidence, for example, is that massive stars tend
to form in clusters, with the most massive in each cluster
often formed last.}.
Thus, there is a kind of a
feedback effect which limits the efficiency of conversion of
the giant molecular
clouds to stars.  Indeed, the very massive stars radiate in
the ultraviolet, which ionizes the gas around them.  These hot,
ionized regions are found
surrounding sites of recent massive star formation.

Third,
while the clouds are dense and cold enough to collapse
gravitationally, it seems that they are supported against collapse
by some combination of turbulence and magnetic fields.
 This means that the rate of star formation can be greatly accelerated
if the cloud is subject to an external perturbation such
as a shock wave.  Indeed, while low mass stars may spontaneously
condense out of the giant molecular clouds, it is widely believed
that the formation of massive stars would be much
rarer in the absence of these external perturbations.

The main source for these external perturbations is believed to be
other recently formed massive
stars\cite{elmegreen-triggered}.  Primarily through
supernovae, but also possibly through their ultraviolet
radiation, very massive stars
form shock waves in the interstellar medium.  While these may
destroy the giant molecular clouds in which they form, the result
seems to be the catalysis of  massive star formation in
nearby giant molecular clouds.

This gives rise to a phenomena which is called self-propagating
star formation\cite{IBMguys}.  As long as there is a
continual supply of giant
molecular clouds, the formation of massive stars can spread through
the disk through a process in which the supernova of a massive
star in one cloud catalyzes the formation of new massive stars in
nearby clouds.  The time scale for this process is the lifetime of
a massive star, which is at most $10^7$ years.

There are several independent pieces of evidence that the
rate of star formation in the disk is governed by feedback
processes occurring at several scales.  The first is simply
the fact that the interstellar medium maintains a configuration
consisting of a number of different phases with approximately
constant proportions of mass and volume.
This is a dynamical
stability, as the presence of the different phases means that the
medium is out of equilibrium.  Moreover, the giant molecular clouds
must be condensing out of the warm gas as a steady rate, as
they are being continually destroyed through the
process of star formation.
Further, the evidence shows that the star formation rate in
our galaxy, and other similar galaxies is now to a good approximation
equal to the average rate over the lifetime of the galaxy, which is
about $10^{10}$ years.
The time scales for the processes involved are small, compared
to this lifetime,
$10^5$ years for the collapse to new stars and $10^7$ years for
the time between formation and supernova of a massive stars;
to maintain this non-equilibrium configuration stably over
so many dynamical times there must be feedback processes
that control the rates of formation of clouds and stars.

Further {\it a priori} evidence for the existence of processes
governing these rates is that the rate by which material is
converted from the interstellar medium into stars, which is
about $3-5$ $M_{solar}$ per year, matches well the rate at
which matter is returned from stars to the medium through
supernova and stellar winds, which is estimated to be
at least $1-2$ $M_{solar}$ per year.  To astrophysical accuracy,
these numbers could be equal, but even if they are not, they are
close enough that some explanation is needed.  Related to this is
the fact that although star formation has been going on for
$10^{10}$ years, it is the case that in the midplane where these
processes take place, fully half of the matter is
presently in gas and dust.

In thinking about these things it is important to emphasize that
the galactic disk seems to be an open system.  Old stars evaporate
off of the midplane at a constant rate, as their encounters with
other stars give them velocities perpendicular to the plane.  Further,
it may be the case that new gas continually or intermittently enters
the system, either by infall onto the disk or by inflow into the
star forming regions from the gaseous disk that seems to extend
quite a bit out of the visible galaxy.

Another kind of evidence that there ought to be mechanisms
that control the rate of star formation in spiral galaxies is that
there are galaxies where this apparently does not happen. Little or
no star formation is taking place in elliptical galaxies; they contain
no dust and what little gas they have is heated to the point where
further star formation seems unlikely.  At the other end of the
scale are the so called star burst galaxies that are forming stars at
rates that are not sustainable for long periods.  Many of
these are small or dwarf galaxies, which seem to be
found in either a star burst mode or in a quiet mode with little
star formation.   The evidence is then that to achieve the steady,
sustainable rates of star formation that are seen in spiral disks
requires a system of a certain size.

All of this evidence suggests that there must be mechanisms that
explain how a spiral disk achieves a steady rate of star formation.
Several hypotheses have been made about such mechanisms.  I
will describe a few of them.

Parravano has proposed a feedback mechanism that regulates
the rate of star formation by controlling the
rate of condensation
of the giant molecular clouds\cite{parravano}.
The idea is that the mechanism
maintains the interstellar medium at the critical temperature
at which the giant molecular clouds may exist in equilibrium
with the warm ambient gas.   This critical temperature depends on
the pressure and other factors such as the amount of dust and
there is evidence that the interstellar media of
a large number of galaxies are near it \cite{parravano}.
A mechanism
that would maintain the medium at this critical point works
as follows.  As the temperature falls below the critical point
giant molecular clouds condense, which leads (in combination
with other factors) to the formation of new massive stars.  The
ultraviolet light from these stars heats the medium, raising
its temperature above the transition.  This cuts off the formation
of new clouds, and hence new stars.  But after about $10^7$ years
this leads to a decrease in the ultraviolet radiation,
leading to the cooling
of the gas below the critical point, and so on.

What is interesting about this mechanism is that it functions on
scales larger than individual clouds, tieing the rate
of star formation
to the rate of cloud condensation.  If there is a mechanism to
guarantee massive star formation given the existence of giant
molecular clouds it can explain why this process may continue
at a steady rate as long as the pressure is sufficient for there to
be a critical temperature.  It is then interesting that the supernovae
can both provide a mechanism for star formation given the
presence of enough molecular clouds, through the self-propagating
star formation, and provide the energy which pressurizes the
interstellar medium.

Let us then assume that there is a mechanism such
as Paravanno proposes that keeps the
medium critical  so new molecular clouds are condensing
from the ambient gas at a steady rate.    There will then be a steady
rate of star formation. We then want to ask more detailed questions
about the geometry of the star forming regions.  This kind of
question was addressed by the Seiden-Schulman-Gerola model.
In this model the process of self-propagating star formation is
described as a percolation process, which is then modeled by a
cellular automata.  The model is simple, in some ways very like
the game of life, put on a rotating lattice.

The model is constructed by dividing the disk into rings, each of
which is divided into a number of cells.  The disks rotate
differentially, at the
same linear speed, in order to match the flat "rotation curves"
that are generally observed in disk galaxies.  The model evolves
in discrete time steps according to simple local rules.
Each time step is about $10^7$ years, which is the typical
time between the birth of a massive star and its destruction
in a supernova.  At each
step each cell may be on or off, which represent whether star
formation is occurring or not.  Each cell also has associated with
it a quantity of gas, which is distributed among two states, which
correspond to the warm ambient gas and the cold clouds.  It is
assumed that in each time step in which star formation does
not occur in a cell, a certain proportion of its gas condenses from
warm to cold clouds.  But during a step in which
star formation occurs, all the
gas is heated and returned to the warm state.

The rule of evolution is then stochastic.  There is a small spontaneous
probability for star formation to occur and an induced probability
which is proportional to the number of neighboring cells in
which star formation occurred in the last time step multiplied by the
amount
of cold gas in the cell.

The model has three parameters:  The radius of the galaxy, which
gives the number of rings, the rotation velocity and the rate at
which cold clouds condense from warm  gas.  The latter gives
a ``refractory period" over which star formation is unlikely to
repeat in the same cell due to there being insufficient cold clouds
for stars to form.  Over a wide range of parameters, the model
seems to show what  might be called self-organized
critical behavior, in which star formation occurs at a steady rate.
In this state, the star forming regions make spiral patterns that
continually form and dissolve.  Furthermore,
given a suitable choice of parameters, these resemble
rather well the patterns seen
in some spiral galaxies.

It is possible to interpret the model in the following way: the
dynamics of the gas is providing a feedback mechanism which is
tuning the system close to the critical point of a percolation problem.
Indeed, one may simplify the model by eliminating the gas.  Then
the induced probability for star formation to occur in a region
is simply proportional to a parameter times the number of neighbors
in which stars are forming.  In this case the system is a directed
percolation problem in $2+1$ dimensions.  There is a
percolation phase transition and to get spiral patterns
that continually form and a constant rate of star formation one must
tune that parameter near the critical point for the transition.
What the gas dynamics seems to do is to eliminate the need for
adjustment of a parameter by keeping the system in the critical
state by a feedback process.

Some astronomers have criticized this model for oversimplifying
the real phenomena and also for being unable to describe certain
kinds of spiral galaxies.  Their criticism is in part correct, but
in a way also misses the point.  It is true that important
phenomena are neglected in this model, for example the
gravitational dynamics of the stars and the medium are
completely ignored.  It seems that in some galaxies this
is justified.  In these, the spiral patterns are seen only in the
distribution of star forming regions, and hence are observed
only in the blue light coming from the bright young stars, and not
in the red light coming from the old stars.  In these the spiral
patterns tend to be fluffy or ``flocculent", and it is these kinds of
patters that seem to be well modeled by the Seiden-Schulman-Gerola
model.

In the older models favored by some astronomers, the opposite
simplification is made.  The gravitational interactions among the
stars are modeled, and the energetics of star formation and
supernova, as well as the processes governing the conversion of
material between stars and the interstellar medium are ignored.
In these models one sees that density waves can be excited in the
distribution of stars in the disk.  These can catalyze star formation,
because  a giant molecular cloud can be perturbed significantly
by falling into the deeper gravitational potential of a passing
density wave.  According to such a model, the spiral patterns
should be seen both in the new stars, tracing the star forming
regions, and in the old stars, tracing the density wave.  Furthermore,
in such models the density waves, and hence the spiral patterns
can show bilateral symmetry, so one can have strongly symmetric
spiral arms, and not just a kind of spiralling fluffy pattern.

Galaxies of this kind, in which the gravitational dynamics seem
important, are seen.  Clearly these are not going to be modeled
by the Seiden-Schulman-Gerola model.  However, the density
wave models have problems of their own.  The density waves
must be excited, either by an outside perturbation such as a passing
galaxy or by an asymmetric field such as might be generated
by the galaxy itself.  Such asymmetric structures are seen, they
are usually in the form of bars.  However, spiral structures
are seen in galaxies that are without bars and are also apparently
isolated far from other galaxies.  In these cases there is a
problem as the density
waves are damped, and will die out after a few rotations.

Clearly what is needed are models that contain both elements.
Although it is not the most elegant possibility, it is hard to avoid
the conclusion that there are
some galaxies in which the spiral structure
traces density waves and there are other galaxies in which the
spiral structure is not traced in the density, and is more a
result of self-propagating star formation near a percolation phase
transition.  This point of view has been advocated by Elmegreen,
who, with Thomasson has constructed such a hybrid
model\cite{elmegreen-model}.  In this
model, the gravitational dynamics of the stars and the energetics
of star formation and the interaction of stars and clouds are included.
This model seems, for appropriate choices of parameters, to be able
to describe either kind of spiral structure.

At the same time, while it describes a wider range of phenomena,
the Elmegreen- Thomasson model requires
that certain parameters that
describe the energy balance between the populations of stars and
clouds be tuned so that a constant amount of energy is maintained
in the system.  This tuning of rates to maintain the energy
balance is presumably accomplished in nature by the kind of
feedback mechanisms that are modeled by the Paravanno
hypothesis and the Seiden-Schulman-Gerola model.  Thus, while
it may be a satisfactory model of spiral structure, the
Elmegreen-Thomasson
model still does not represent a complete model
of the energetics of
the disks of spiral galaxies.

But my point here is not to debate whose model is better but
to make the point that it may be useful to describe the
disks of spiral galaxies as self-organized critical systems.
Let me then end this section by summarizing the reasons why
it seems reasonable to think of the disks of spiral galaxies as
self-organized critical systems.

That they
are critical is to be seen from:

\begin{itemize}
\item The simultaneous presence of several
different phases in the interstellar medium with very different
densities, temperatures and compositions, again over very long
time scales.

\item The evidence of Paravanno that
many galaxies are near the transition point for simultaneous
existence of warm gas and cold clouds.

\item The suggestions that the distribution of densities
in the cold molecular clouds is scale invariant.

\item The apparent  long ranged order in the spiral structure, which,
together with the mechanism of self-propagating star formation
suggests a percolation system near a critical point.
\end{itemize}
The evidence that they are self-organized comes first of all from
the evident fact that as galaxies are isolated, any critical behavior
that is widely seen must be arrived at spontaneously, without
the need for tuning of external parameters.  Besides this,
there is evidence from,
\begin{itemize}
\item Constant star formation rates, over time scales very long
compared
to the dynamical time scale.

\item An approximate balance
between the rates of flow of matter in each direction between stars
and the medium in the disk, despite the possibility of loss of stars
by evaporation to the halo and inflow and infall of gas to the disk.

\item  The success of the hypotheses of Paravanno, Seiden, Schulman
and Gerola and others that there are feedback mechanisms which
maintain the gas in the disk in a critical state with a constant
rate of formation of cold clouds, which matches their rate of
destruction.
\end{itemize}

\section{What is the large scale organization of the universe?}

Probably the key cosmological problem at present is that of the
formation of the galaxies and their large scale organization.
The amount of
data we have about the history and organization of the
universe on scales larger than galaxies is increasing quickly;
and the theories have consequently been evolving very
rapidly in this domain.

Given the apparent usefulness of conceiving of a galaxy
as a self-organized non-equilibrium system, it is natural
to ask if new concepts from non-equilibrium statistical
physics such as self-organized criticality might be useful for
understanding
how structures on still larger scales emerged.  There are three
reasons, {\it a priori} for imagining that this might be the
case.  First, the processes that formed the galaxies and their
large scale organization occurred at earlier times when the
universe was on average denser and hotter.  It is then natural
to ask if non-equilibrium processes such as those we
see  dominating the process of star formation might have
played a role in some denser era in the formation of galaxies.
To put this another way, we now understand galaxies to be
dynamical systems, in which supernova and other
energetic processes play a dominant role.  It is then natural
to ask whether such processes might have played a role in
their formulation.
 Second, there are senses in which
the distribution of galaxies and clouds of gas are approximately
scale invariant, which may suggest the study of galaxy formation
as an example of a critical system.
Third, there is a sense in which all gravitationally bound
systems are intrinsically out of thermal equilibrium.

I would like to briefly expand on this last point.
While  gravitationally bound systems may spend
long periods of time in quasi equilibrium configurations,
they do not reach true equilibrium states, characterized
by maximal entropy\footnote{There actually is available
an equilibrium state for any isolated gravitationally bound
system, it is the black hole containing the total mass and
angular momentum of the system.  Fortunately, the time
required for most astrophysical systems to reach this
state is much larger than the Hubble time.}.
The reason is that they have
practically inexhaustible sources of energy, coming from
gravitational binding energy of subsystems.  A subsystem
can always become more deeply gravitationally bound,
freeing energy to other parts of the system.  A consequence
of this is that all large gravitationally bound systems
are characterized by a flow of energy at some rate from
gravitational energy to heat or kinetic energy.  The question
is only the rate.  When coupled with another source of
energy, nuclear energy, gravitationally bound systems such
as galaxies can maintain significant flows of energy for
cosmological time scales.

It is significant that what characterizes self-organized
critical systems is that they are kept out of equilibrium
by steady flows of energy through them from a source
to a sink.  Large gravitationally bound systems do this
naturally.  It is then interesting to speculate that all
large gravitationally bound systems may, to one extent
or another, be thought of as self-organized critical systems.
This description apparently is suitable in the case of spiral
galaxies, it is then interesting to ask if the flows of energy
in other systems and on other scales is significant enough
to play a role through mechanisms of self-organization.

The evidence we have presently for the large scale organization
of the universe comes from many sources.  The most important
methods have been
1)  catalogues of galaxy redshifts; 2)  absorption lines
in quasarspectra, 3)  the cosmic black body radiation, 4)  studies of
the distribution of hot ionized gas in
clusters of galaxies, by measurements
of the x-rays they give off and 5) measurements of
large scale velocity flows, by careful combinations of
distance and redshift measurements.   Together these
give a detailed picture of the organization of matter in
the universe, and the amount of data available is expected
to increase dramatically over the next years.  A theory
of cosmology must account
for all of
these data by a detailed description of the history of the universe
that begins perhaps $10^{-43}$ after ``the big bang" and runs
to the present.  It is a tall order, and it must be said that the
existing hypotheses do not do badly at the present time.
But there are issues that suggest that the present picture is
incomplete; I sketch here a few of them.

\subsection{Dark matter and the issue of $\Omega$}

Any understanding of the large scale organization of matter
in the universe must take into account the evidence that
at least eighty percent of it is not visible.   The strongest
evidence for this comes from the rotation curves of
galaxies, which leads to the conclusion that most
galaxies are surrounded by large halos of non-luminous
matter, with between five to ten times as much mass as
is present in visible stars, gas and dust\cite{peebles-book}.
In units
of $\Omega$, where $\Omega=1$ would be exactly enough
matter to close the universe, the visible matter in galaxies
is about $\Omega_{observed} =.01$, whereas the total gravitational
mass in galaxies
is roughly ten times larger.

Other evidence comes from careful studies of clusters of
galaxies.  Measurements of X-rays from large clouds of
ionized hydrogen surrounding the galaxies lead to a conclusion
that there is no more than about ten times more gravitating matter
than is contained in the observed baryons.  This, together
with the bounds coming from nucleosynethesis, which
is $\Omega_B h_{50}^2 = 0.05 \pm 0.01$, leads to the
conclusion that
that $\Omega = .1 - .2 $ \cite{Omega-WNE}.

The question is then whether there might be still additional
non-luminous matter, clustered on still larger scales, that
could increase
$\Omega$, perhaps to unity.  While the evidence for a low
value of $\Omega$ in the range $.1-.2$ seems to be
increasing\cite{Omega-WNE},
we may note that there are observations of large scale flows
of matter that, given certain theoretical assumptions, point
to a larger value\cite{structuremodels}.
A number of other observational
issues bear on this
question including the value of the Hubble constant,
the question of the mass of the neutrino,  the
age of the oldest stars in globular clusters, and
the abundances of rare primordial elements.  It seems likely
that there will be significant progress on all of these questions,
so that we may hope within a decade or two for a sharp resolution
of the value of $\Omega$.

There is a strong theoretical reason for a value of $\Omega =1$,
which is that it seems to be required by all natural inflationary
scenarios.  It is possible to invent inflationary models for which
$\Omega < 1$, but these require an additional
tunings of a certain parameter\cite{openinflation}.
Theorists may disagree on the extent to which this is a cause for
worry, as there are already at least two fine tunings that must
be done for any inflationary scenario to work, and to yield
a reasonable spectrum of fluctuations, first of the
cosmological constant and second of the self-coupling of the
``inflaton" field.  This is not to say that fine tuning is not a problem,
but only that if inflation is to be in the end accepted we must
uncover a natural mechanism to accomplish these fine tunings;
if such a mechanism is discovered it may as well be able to
fine tune the inflationary mechanism so that $\Omega < 1$.

If $\Omega$ does fall in the range $.1-.2$ favored by most current
observations, it may free theory from having to provide an exotic
non-baryonic  particle for the dark matter.   Given the apparent
failure of pure hot dark matter models, we know the non-baryonic
dark matter cannot be only massive
neutrinos;  so any theory that demands $\Omega$ to be equal
to unity requires that we postulate that the universe is dominated
by a kind of matter for which we have no observational
evidence.

On the other hand, if $\Omega \neq 1$ then the universe
has an intrinsic scale written into it's initial conditions,
which is greater than its present age.  Assuming that the
initial conditions are set at some early time by the
action of physical processes involving quantum gravity
or grand unification, leads then to
a puzzle, for we must ask how physical processes
involving time scales of $10^{-43}$ seconds could be
fine tuned in a way that implicitly involves a time
scale of $10^{17}$ seconds.

\subsection{Quasar absorption lines and the universe at earlier
 times}

A window into the distribution of matter in the universe of
increasing importance  is the analysis of the absorption lines of
quasars.  Many quasar, have  redshifts in the range of $z=2-5$, and
were thus active when the universe was significantly smaller.
It turns out that whenever the light from
a quasar passed through a sizable enough cloud of gas on its
way to us we see absorption lines at the appropriate redshift.
Most of these lines are due to the Lyman alpha transition in
hydrogen, and some are produced by heavy elements such as
carbon and magnesium.

More than $150$ quasar
spectra have been studied, and each of them
contains on the order of $100$ lines, so that there are
reasonable
statistics about the distribution of clouds of gas between
them and here.

The basic results seen in these observations are the
following\cite{qsoabsorb},
\begin{itemize}
\item There is little or no unionized hydrogen between the clouds.
For example, at a redshift of $2.26$, the ratio of unionized
hydrogen seen to the average matter density is less than
$10^{18}$ \cite{qsoabsorb}.
This most likely means that the intergalactic medium is ionized,
up to at least a red shift of $z=5$.
The source of the energy
to ionized the medium is unknown; this is itself an important
problem.  Possible candidates are the quasars themselves,
an early generation of supernovas or massive stars. There
are also exotic possibilities, such as the decay of massive
neutrinos.

\item From the Lyman alpha absorption lines one may measure the
column density of neutral hydrogen in each cloud, which is the
number of atoms per square centimeter
in the line of sight of the quasar through it.
Remarkably, over a range
of at least nine orders of magnitude, from
$10^{13}$  to $10^{21}$  $atoms/cm^2$,
the distribution of clouds at a given redshift satisfy a power law
distribution in column density $\sigma$;
\f
n(\sigma ) \approx \sigma^{-\gamma}
\ff
with $\gamma =1.67\pm .19 $ \cite{qsoabsorb}.

At the high end, the column densities are comparable with
those through the central region of the disk of a spiral
galaxy.  It is intriguing that these are seen to fit into a
single power law with much more diffuse column densities.

Because we are seeing through a random line of sight through
each cloud,
the distribution of column densities may be a combination
of two factors, the distribution of densities within a given cloud
and the distribution of masses of the clouds themselves.  One may
make a number of hypotheses about both.  However, whatever
combination of these factors determines the power $\gamma$,
the fact that there is one power that ranges from the densities
of galaxies down suggests that one mechanism must be responsible
for the formation of the galaxies and the clouds seen in
the absorption lines.  This is particularly impressive as there
are so many orders of magnitude
involved\footnote{There is also the possibility of a break
in the distribution, so that the distribution has slightly different
powers at high and low column densities\cite{jane-break}.}

\item One hypothesis that may be explored is that the galaxies are
surrounded by large diffuse clouds of gas, that are in approximate
hydrostatic equilibrium, and so have densities that fall off
like $r^{-2}$ as we go from the center.   There is increasing
evidence for such a picture in the study of correlations between
the denser
absorption lines and actual galaxies near to the line of sight
of a quasar\cite{qsocorrelations}.

Very recent observations suggest find that, at least
for low redshifts, if a galaxy is within
$40 kpc$ of the line of sight there is almost always an
absorption line in hydrogen with a column density of
at least $10^{15} / atoms/cm^2$
and vice versa\cite{qsocorrelations}.  This suggests a
picture in which many galaxies
are surrounded by spherical
clouds of hydrogen and other gases which
extend out to at least $40 kpc$.  These clouds, are often seen
also in carbon and magnesium, so that it appears that they
have been enriched by the action of supernovae. It is then
very interesting to know whether this enrichment came from
supernovas at an earlier time, took place during the formation
of the galaxy itself, or, on the other hand, is the result of outflow
from the galaxies themselves.

It is then interesting to try to imagine that these clouds and
the galaxies they contain are single systems, with significant
exchanges of matter between then, perhaps in both directions.
One may wonder, for example, whether the observed constant star
formation rates of spiral galaxies are related to the rates at which
gas falls from the surrounding clouds onto the disk.

\item Finally, the quasar absorption spectra give very good
probes of the distribution of matter at high redshift.  One
intriguing result is that at very high redshift $z>4$ there are
about four times more of the densest absorption lines than
would be given by the present day galaxies.  The interpretation
of this is problematic; it may be that many clouds never
formed into galaxies, or it may be that the clouds have
contracted significantly since then.
\end{itemize}

However, we must keep in mind another interesting thing,
which is that
there is evidence that the properties of large
galaxies have not changed very much since redshifts of
$2-3$, which is on the order of ten billion
years\cite{noevolution}.  Before that
time, energetic processes, such as those that fuel quasars,
were much more common then they are presently, however there
seems to be a sharp decrease in the numbers of quasars seen after
red shifts around $2$,
suggesting that large normal galaxies have since that time
established a kind of
equilibrium\footnote{However, it should also be mentioned that
the observations indicate that the much smaller ``dwarf"
galaxies have evolved a great deal since redshifts of $2$,
there seem to have at that time been more of them than there
are now, especially the ``blue" ones, in which a lot of
star formation is going on\cite{dwarfs}.}.
The evidence we mentioned
above agrees with this picture, suggesting that normal spiral
galaxies have a constant rate of star formation over most of
the time since their formation.  This, together with the evidence
I summarized above,
suggests that it might be fruitful to understand the
galaxies and their surrounding gas clouds as single
stable far from equilibrium systems.

\subsection{The issue of homogeneity on very large scales}

There is a final issue I should mention,
which has been the subjection of discussion
among statistical physicists.  This is the question of the large
scale homogeneity of the universe.

Since Einstein and DeSitter, cosmological models have always
been based on the Cosmological Principle, which assumes that
we live in a typical place in the universe.   It is also observed
that to very high precision, the universe is isotropic to a very
good approximation.  This can be seen in the $COBE$ radiation,
which is isotropic up to a part in $10^5$.  As the radiation has
passed through the gravitational potential of matter on its way
here, this puts limits on the anisotropy of the distribution of
matter from redshifts of $1000$ to the present.  Counts of
galaxies, or radio sources also show impressive evidence of
isotropy\cite{peebles-book}.

If our view of the universe was
perfectly isotropic, and it were so, by the
Cosmological principle, around every point, then we would have
to conclude that it was perfectly homogeneous.   The difficulty
is that it is neither
perfectly homogeneous nor perfectly isotropic, which
raises the issue of how it is to be described.

The simplest assumption is that the departures from homogeneity
decrease at large scales, so that there is some
scale $\lambda_h$
above which the universe may be satisfactorily described as
homogeneous\footnote{At least up to some larger scale, we may note
that no cosmological observation is able to constrain the
homogeneity of the universe on scales larger than the distance
to our horizon, so that it is perfectly possible that the universe
is very inhomogeneous on some much larger scale.  This possibility
is taken advantage of in the inflationary models, which describe
the universe as a single bubble that inflated.  The bubble does
have walls, even if we can't see them.}.  This assumption is
usually made by astronomers, and so far there is no evidence
against it.

The difficulty is that the large scale surveys of the galaxies,
which map the distribution of matter, show so far structures
that are as large as the scales of the
surveys\cite{structure}.  Furthermore, at
least up to the scale of clusters of galaxies, the distribution of
matter is
approximately scale invariant.  This means that
one of two things must happen.  As the surveys increase in
depth, the scale $\lambda_h$ must be discovered, or
structures must continue to be
found on every scale up to the horizon.
It has been thus suggested that perhaps the standard assumption
is wrong, and the universe has a fractal (or multifractal)
structure on all scales up to the horizon\cite{}.  The difficulty
with this picture is that such a distribution should also agree
with the isotropy seen in both the counts of galaxies and
radio sources and in the microwave background, as well
as with the Cosmological Principle\cite{}.
The question is whether
there can be a distribution that shows inhomogeneities on
arbitrarily large scales that is in
agreement with this.

A related question is how to
describe a universe that is inhomogeneous over a large
range of scales in general relativity.  Clearly it will not
do to work solution by solution, what is needed is something
like a renormalization group treatement, that lets us
think in terms of coupling between modes on different
scales.  The tricky thing is how to to do this in a way
that is generally covariant, since the metric that measures
scales is dynamical.  A very interesting step in this direction
has been taken by Carfora and
Piotrkowska\cite{MauroKamilla}.   Even if there is a
scale above which the universe is homogeneous to a good
approximation, there are corrections to the equations that
describe the expansion of the universe coming from averaging
over the fluctuations at smaller scales.  An important, and presently
unresolved question, is to determine if these corrections
may lead to significant modifications in the age of the
universe\cite{MauroKamilla}.

\section{The problem of the origin of the
large scale structure}

We have been discussing the evidence that tells us how
the universe is organized on large scales.  Now I would
like to turn to the question of what is understood
about how that structure has arisen.

The first thing that must be said is that astronomers have
developed numerical models of the evolution of structure
in the universe that seem to  go quite far
towards explaining features of the observed distribution of
galaxies.
I would like to begin
this discussion by summarizing
how these models
work\cite{structuremodels,peebles-book,OstGen}.

The models take as
inputs certain assumptions about the
conditions of the universe at decoupling.  These begin with
a specification of the basic cosmological parameters, such
as $\Omega$, the value of the
cosmological constant and the amount of
dark and baryonic matter
present.
Because of the isotropy of the present universe, and the fact that
it works so successfully, the universe is always assumed to be
homogeneous and isotropic, with an initial spectrum of perturbations
whose amplitudes are small (on the order of $10^{-5}$) on all
scales.

To this picture one then must add several assumptions.
First, one must choose between two general
types of initial perturbations.  Adiabatic perturbations are those
for which the baryonic and photon densities fluctuate together, so
that the observed temperature fluctuations observed in the
COBE signal trace also fluctuations in the density of baryons.
Another choice is to take what are called ``isocurvature"
perturbations, in which there may be larger fluctuations in the
density of baryons, which are, however, not reflected in the
distribution of temperatures, because the distribution of thermal
photons does not trace the distribution of matter.  The first is
better studied, but both are reasonable possibilities.

A very important assumption that must be made is the spectrum
of initial fluctuations.  The assumption that is most often made
is that the initial spectrum of fluctuations is approximately
scale invariant, this is the simplest possibility and was proposed
some time ago by  Harrison and Zeldovich.  It is also what is
predicted by inflation.  The amplitude of the spectrum may then
be normalized by the COBE measurement.

In the near future the measurements of the black body spectrum
are expected to be very much improved, so that the initial
spectrum of fluctuations will, in the adiabatic case, be largely
constrained by observation.  Of course, this will not constrain
the isocurvature models as much, as by assumption they take the
initial perturbations in the baryons to be decoupled from those
of the photons.

The last assumption that is made in the construction of these
models is the type of dark matter present.  These may
be of several kinds: dark matter may be hot or cold, depending
on whether their masses are small or large compared to the
cosmic background temperature, it may also be baryonic, in
the case it consists, for example of black holes, or non-baryonic,
as in the case of neutrinos or hypothesized particles such
as axions.

Given these choices, the numerical simulations have been able
to show how the perturbations grow, leading to the structures
we see today\cite{structuremodels}.
While there are important differences between
the models based on different assumptions,
a variety of assumptions are known to lead to
structures very much like those we see today.  Very roughly,
in all of them perturbations grow through a long linear phase, from
their initially small values to values of order one.  After that,
non-linear processes involving both gravitational binding and
hydrodynamics effects take over, leading more or less quickly
to the formation of galaxies and clusters of galaxies.

It must be emphasized that it is nontrivial that the models work
at all, given the simplicity of the assumptions made.  Given that the
spectrum of perturbations at decoupling
is constrained, in the adiabatic
case, to such a small value by the COBE data, and given that the
age of the universe is also constrained, to within a factor of two,
it might very well have been the case that structure forms at
too slow of a rate to explain what is seen at the present time.

There are, however, a number of places in the picture in which
complementary or alternative points of view may play a
role.   These include particularly the role of non-linear
processes in structure formation.

\subsection{Understanding the non-linear
stages of galaxy formation}

 According to the standard models of structure formation,
once the perturbations in the distribution of mass and
baryons become of order one, non-linear processes take over,
leading to the formation of the present day structures.  While
there is a good analytic description of the linear regime, there
is no correspondingly successful treatment of the non-linear
regime besides the large scale numerical simulations.

There are several possible
indications that self-organized critical
phenomena may play some role in this non-linear regime.
\begin{itemize}
\item The structures which are formed are scale invariant, and
governed by power law distributions at least up to the
scales of clusters of galaxies.  Furthermore, as I mentioned
above, the structure of
clouds of gas, back even to red shifts of $4-5$ follows
a power law over $10$ orders of magnitude, as seen in the
distribution of quasar absorption lines.  Thus, irrespective
of the question of the large scale organization, it is clear
that the distribution of galaxies and gas may be characterized
as fractal over many orders of magnitude.
\item There are suggestions that several features of the final
distribution of galaxies and mass may be independent of
the detailed assumptions that go into the large scale simulations.
These may include the powers that govern the distributions of
galaxies.  If so, this suggests that there may be simpler statistical
arguments for some features of the observed distributions.
\item There is a rather simple model of hierarchical
structure formation
due to gravitational binding in an expanding universe that does
agree to some extent with both the observed distributions and the
results of the numerical simulations. This is the Press-Schecter
formulation \cite{pressschecter}, which I would like
to briefly describe
\end{itemize}
In a very interesting paper, Press and Schecter described both
analytic and computer models of
a collection of gravitating particles in an expanding
universe.  The particles originally have equal masses and
are randomly distributed.  At certain intervals,  as the universe
expands, a test is applied to identify clusters that are gravitationally
bound.  In their computer simulations Press and Schecter
considered spherically regions which were overdense by a
factor of 10 to be bound.  Those clusters that are bound at
each step are replaced by single particles with a mass
which is the sum of the masses of its members.

This processes is iterated and it is found that after a time an
approximately
scale invariant distribution of masses  develops which
has the form
\f
n(M ) \approx {1 \over M^{1.5} }
\ff
for small masses, times a high mass cutoff $e^{-kM/R^2 }$
that scales as $R$, the scale factor of the universe.

Press and Schecter found that the same approximate
scale invariant distribution resulted from their model,
given different kinds of initial distributions of the
particles.  They also gave a simple analytic
derivation of the scaling law.  Finally, they
were able to compare the predictions of this
model with observation and they found
that the distribution of luminosities
(which  scale with mass) of galaxies in the Como cluster
scale with an approximate power law.  Since that work
was done,  both observations and numerical simulations
of the distribution of galaxies in clusters has tended to
support this simple picture\cite{support-ps}.

We may note that in the formation of an
apparently universal  scale invariant
distribution from different initial conditions, the
Press-Schecter model might be described as a very simple
kind of self-organized critical system.

\subsection{Possibilities for early structure formation}

Finally, even though the simulations of galaxy formation based on
the standard dark matter scenarios do seem to work reasonably
well, there is still the possibility that they are based on assumptions
that may turn out to be incorrect.  Especially given that some
features of the observed distribution of galaxies may be produced
by non-linear effects that wash out some of the information about
the initial conditions, we must keep open the possibility that
more detailed observations, especially at higher redshifts, may turn
out to be inconsistent with these models.  It may then be useful
to consider other kinds of models which may account for the
observed structure\footnote{Many suggestions have
been proposed that depart in small or large ways from the
standard structure formation scenario I sketched here.
Several involve explosions or other energetic events in
the early universe\cite{oc,ikeuchi,osetal}.  Others interesting
proposals
involve a low density, $\Omega \approx .1-.2$ universe
\cite{OstGen,Peeblesbaryonic,hogan}.}
While this might be considered a higher
risk activity than the others on my list, it may be motivated
by consideration of the fact that there is a certain lack
of economy in the assumptions that must be made in the standard
models.  At present, the nature and properties of both the dark
matter and the initial perturbations are essentially ad hoc, and
can be manipulated to yield results consistent with observations.
It would certainly be more elegant to have a theory in which
there was not so much room to introduce ad hoc elements.

One might then dream that a scenario for cosmology could be
made to work in which
nonlinear processes play a role much earlier in the history
of the universe, acting near or just after decoupling to produce
the  spectrum of fluctuations that become
the large scale
structure\footnote{Two very interesting attempts
to model structure formation in the distribution of galaxies
are by Chen and Bak\cite{ChenBak} and
Schulman and Seiden\cite{SchulmanSeiden}.}.
In such a picture, the slow growth of primordial fluctuations
after decoupling
would be replaced by a picture in which non-equilibrium processes
act at very high redshifts of 500-50 to produce a spectrum of
fluctuations in the distributions of matter and baryons
that might be largely independent of whatever
initial perturbations are present at decoupling.

We may note
that the fact that the isocurvature models are consistent with
present knowledge means that it may be that the perturbations
seen in the black body radiation do not trace the perturbations
in the matter (although there are limits based on
the motion of the light through the inhomogeneous gravitational
fields of matter since decoupling.)

Such a scenario could take advantage of the fact that at redshifts
of around $100-200$ the conditions of the universe as a whole
are not that dissimilar, in density and temperature, from those
which characterizes the interstellar medium of the disks of spirals
galaxies.  It is then possible that non-linear processes
that are analogous to those that are
responsible for the spiral structure in galaxies might
act to form structure at these earlier times.  The amount
of time that such processes would have to act is limited
by the expansion speed at that time to at most several
hundred million years.  But this is one to two orders of magnitude
longer than the lifetimes of the massive stars, making it possible
that processes in which massive stars are formed and inject
a great deal of energy into the medium could produce significant
structure during this time.

There are in fact some
reasons to believe that there was an era of star formation
before the formation of the present day galaxies, coming from
the need to explain both the fact that some enrichment is seen
even in very old clouds of gas and the fact that the intergalactic
medium is ionized back to at least  redshifts of around $5$.
At the same time, such a scenario would have to be limited by the
requirement that not too many heavier elements were
produced\cite{carr-dark}.

One may also try to understand if, in the context of
such a scenario, it is possible if the dark matter could
be formed as a consequence of such early processes of
structure formation, rather than having to be posited
independently.  One way this might work is if a very
early era of
star formation processes produced large numbers of neutron
stars or black holes, which made up some or all of the
dark matter that then dominates the structure and formation of
the galaxies at later redshifts from about $20$ to the present.
A dark matter scenario in which the non-luminous matter
consists of black holes which are  formed in the same processes
that make the
the galaxies and stars might be more parsimonious than
the standard scenarios in which the dark matter is put in by hand
to account for the structure formation.

At the same time, the possibilities for such early structure
formation processes are
constrained from several sides, including limits on the numbers
of black holes coming from MACHO searches and other
observations.

\section{The problem of the parameters in particle physics and
cosmology}

In the introduction I stressed that many of the key problems in
cosmology rest on problems of fine tuning involving the
parameters of particle physics and cosmology.  It is indeed,
not an exaggeration to say that the fact that we live in a world
which is large, complex, out of thermal equilibrium and full of
a large variety of phenomena is a consequence of the parameters
being tuned to special values.  There are two kinds of such
fine tuning problems.  The first involve issues of hierarchies, in
which parameters have improbably small values, such as in the
case of the values of the proton or electron mass, in Planck units.
The other class involves cases in which structures of a certain
kind would not exist if a parameter were to take values different
from its present ones, by less than an order of magnitude.  Examples
of this are the proton-neutron mass difference, the electron mass,
or the strength of the fundamental electric charge: increases in
any of these separately, by factors less than ten would result in
a world with no nuclear bound states, and hence no nuclear or
atomic physics.

There are two responses that have traditionally been made to
the problem of the values of the parameters of particle
physics, in the light of this situation.  The first is to hold to the
faith that there is a unique fundamental theory that
after a pattern of spontaneous symmetry breaking and, perhaps,
dimensional reduction, will have a ground state whose low
energy excitations will match the pattern of elementary particles
and forces that we see.

As the existence of such a theory has been taken to be almost
axiomatic by many theoretical physicists, let me spend a moment
to suggest its likelihood is not so obvious.  First, there is no
evidence for the existence of such a theory, at least at the
perturbative level.  In the last ten years we have learned that
there are very large numbers of perturbative string theories,
which give equally consistent unifications of
the strong weak and electromagnetic interactions with
gravity, but in different dimensions, with different low energy
physics.  It may be that there is one
non-perturbative string theory and these perturbative theories
are all descriptions of excitations of its many ground states.
But there seems, at this point, little evidence for
this\footnote{There is recent evidence that the moduli
spaces associated with the diffferent Kalabi-Yau
compactifications may be connected to each other through
singular configuations that may represent critical points
in the parameter space where certain fields condense\cite{andy}.
It is
then possible that there is a single non-perturbative
ground state in which the quantum state is spread out
over this single extended moduli space.  But, it is also
presently a possibility that what is being described is a
very large family of degenerate ground states, which
are able to tunnel to each other by going through the
singular configurations.}.  Instead,
it may be observed that there seems to be a logic under which,
the more disparate fields are incorporated into a unification by
a gauge symmetry, the more it is the case that the properties of the
low energy excitations depend on a choice of the ground state of the
system.  Thus, in theories which incorporate the Higgs mechanism,
the masses of the low lying states depend on coupling to a
condensate.  If there are many degenerate or nearly degenerate
ground states, with different properties, then it may be said that
the masses and couplings of the light particles are determined
cosmologically, as the ground state may depend on the history
or configuration of the universe as a whole.

Thus, in a certain
sense the  assumption that the properties
of the elementary particles are independent of the state and history
of the universe as a whole is breaking down.    To the extent
that this happens,
elementary particle physics and cosmology become interwoven, and
the Newtonian conception that a  particle in a universe that
contained it alone would be just like a particle in our universe
becomes untenable.

Certainly the inflationary models work in this way, as the spectrum
of light particles is different before and after the phase transition
that simultaneously determines the large scale properties of the
universe.  This also, I would argue
is one lesson we have learned from
string theory in the last ten years; whether or not there is a
nonperturbative string theory whose vacuum states they describe,
the fact of these many different perturbative theories means that
consistency alone does likely not govern the choice of the
phenomenology of the light particles.

If the standard model of particle physics is not to follow uniquely
from demanding only consistency, there must be another kind of
principle which picks out which, of the many equally consistent
worlds, is the one we find ourselves in.  Because of the coupling
between the selection of a ground state and the history of the
universe, this means that the hard questions in elementary
particle physics are likely closely connected with the hard
questions
in cosmology.  It is then remarkable that in both cases these
hard problems involve understanding unnatural choices of
the values of parameters.

The second response that has been given to this situation is
the anthropic principle.  This states (in what is called its weak
form) that the choices of parameters that lead to
our world may be picked out  by noticing that it is among a
rather narrow range in which intelligent life can exist.

Now, as stated this is undoubtably true, indeed, it is an aspect
of the fact I have already stressed, which is that with most
choices of parameters a world would not have the complexity
of ours.  The question is whether this observation can be made
into an explanatory principle.   Rather then deal with this
philosophical question at length (again, this is done elsewhere),
I would like here only to ask if it is possible to do better.  That is,
is it possible that there might be a mechanism that could explain
how the parameters were chosen that accounts for the fact that
the actual values selected
lead to a world with the complexity of ours.

I know of one such theory, that does seem to yield non-trivial
testable predictions. I will briefly describe it here, for more
details the reader may consult
references \cite{evolution1,evolution2,book}.

\subsection{Cosmological natural selection}

This theory comes from two simple conjectures about quantum
gravity, neither of which is really new.  The first is that there
are no final singularities in nature, instead, due to
quantum effects that are ignored in the singularity theorems,
singularities inside
of black holes, and final singularities of cosmological spacetimes
are replaced by ``bounces" as a result of  which the collapsing matter
reverses its collapse and begins to expand again.  This is
an old idea that goes back at least to Lemitre's ``Phoenix
universe" and has been discussed by
Wheeler\cite{wheeler}
and others.  Recently, plausible scenarios for how this might
occur have been discussed in the context of
string theory\cite{martinec}
and inflationary models\cite{gott-talk}.
However, to get definite physical
predictions, as I will show we need know nothing about how this
happens
except simply that each black hole and cosmological singularity
turns, one for one, into a new expanding region of space and time.

The second conjecture I will make is that when this happens the
parameters that govern the low energy physics and large scale
cosmology of the new region differ from the parameters of the
one in which the collapse took place by small random fluctuations.
This is also an old idea, which was proposed, in the case of
cosmological singularities by John Wheeler, who called it
``the reprocessing of the universe."  I need to add to it only
the assumption that the changes are small.  Of course, I will have
to say what I mean by small, I will do this in a moment.

There are also plausible homes for such an idea in grand unified
theories or string theories, as in each case there are large
families of vacua, which correspond to different compactifications
and symmetry breakings.  It is quite plausible that a violent,
Planck scale event like the bounce may force the vacuum to
jump or tunnel from one ground state into a nearby one, leading,
after the region has expanded into a large universe,
to a small change in the parameters of low energy physics.

However, again, while it may be important to develop such theories,
the predictions of this theory are independent of the details.

Our universe has at least $10^{18}$ black holes in it, so that
given these assumptions we are dealing with a universe with an
enormous number of regions, in which we find a distribution of
different parameters of low energy physics and cosmology.

However,
given only these two postulates,  we may make non-trivial
predictions about the parameters that characterize our world
if we add only one more assumption, which is the
Copernican postulate that our world must be a typical member of
this ensemble.  We can then make predictions about our
world if there are  statistical predictions that can be made
about the properties
of randomly chosen members of this ensemble.

We can do this because this theory is isomorphic to models
of biological evolution, in which natural selection is described
in terms of the evolution of probability distributions on
fitness landscapes.
As a result there is a natural mechanism of cosmological self-
organization,
that is formally analogous to biological natural selection.

It goes like this.  We may consider the space of parameters of low
energy
physics to be analogous to the space of genes.  On this space there
is a ``fitness" function, which is the average number of black holes
produced by a region of the universe that expands from a bounce.
Now, just like the fitness functions of biology, this function is
strongly variable, as I said in our universe it is quite large, and
there are simple astrophysical arguments that tell us that with many
values of the parameters it will be much smaller.

The reason the fitness function is strongly variable is worth
mentioning: it is that
it is not easy to make a black hole.  In our universe,
a black hole can only be made if a large amount of matter can
be compressed into a very small space, and for this to happen
there must be rather special circumstances.  The fact that this
happens at least once a century in each galaxy of our universe
may be said to be due to the fact I described in section II
which is that the spiral galaxies are in critical states and
so maintain constant rates of star formation over cosmological
time scales.  Furthermore, the spectrum of masses of stars produced
is power law, so that significant numbers of stars are made which
are larger than the minimum size by enough of a factor that they
can collapse to a black hole even after they supernova and return
most of their mass, and sizable amounts of energy to the interstellar
medium.  For the galaxy to be in the critical state it must be the
case, as I mentioned that the rather complicated cooling mechanisms
which make possible
the giant molecular clouds exist.  But this requires that the universe
be chemically complex.  In short, to a first approximation at least,
our universe can overcome the barriers to formation of black holes
efficiently because it is chemically complex.

But with this theory we may turn this around and postulate that our
universe have the improbable values of the parameters that are
necessary
for such complexity because this leads to one way to maximize
the fitness function, and so make many black holes.

I will not go into details about the statistical arguments, as they
are the same as those that are well known to people who work on
theories of self-organization.  Basically, given the rules as I have
introduced them, the probability distribution for the ensemble of
universes is peaked around local maximum of the number of black
holes.
This means that if our universe is typical, it must have parameters
that are near a local extrema of the fitness function.

This leads to definite predictions about astrophysics, because it has
a simple consequence: all small changes in the parameters from their
present
values should lead to universes that make less black holes than ours.
Thus, the theory is eminently falsifiable; all that would be required
to
kill it is to find one parameter of the standard models of physics
and cosmology, a small variation of which would lead to a large
increase
in the number of black holes produced.  Given that there are on the
order of twenty such parameters, and each may be increased or
decreased,
this gives at least $40$ chances to kill this theory.

After several years of trying, I have not found a definite
counterexample
to this prediction. Unfortunately, with some
exceptions\cite{evolution1,evolution2} every
argument for a change of a parameter going one way or another
tends to
come face to face with some unsolved problem in astronomy.  Two
examples
will suffice to explain this general situation.

Given the fact that the chemistry of ``metals" (astronomers call
anything
heavier than lithium a ``metal"), and in particular processes involving
carbon and oxygen, seem to play a crucial role in cooling the
giant molecular clouds to the point at which massive stars may be
formed, it is natural to argue that if the parameters are changed
so that such elements are unstable many less massive stars, and
hence many less black holes would be made.

The difficulty with this argument is that it is likely that some amount
of star formation did take place early, before these elements were
created, because there must have been early generations of star
formation
to get the process started.  And at least some of those stars must
have
been massive enough to supernova, otherwise carbon would never
have
been found outside of stars.  The question is then how many
massive stars are made, in the absence
of ``metals", compared to how many are now.  Unfortunately, this is
unknown, as all the massive stars made early have by now long been
turned into neutron stars or black holes.  But it is possible that
this question may be answered by future developments in
astronomy.

Without metals, star formation may be primarily a fragmentation
process\cite{rees-fragment}, that might be
modeled fairly simply.  It is also
not impossible
that the power law distribution of masses produced presently by
galaxies can be understood in terms of a description of the spiral
disks as self-organized critical systems.  It is clear in general
that the question of the distribution of stellar masses produced,
either presently or primordially, is a problem in statistical physics.
Of course, if the theory I described here is true, it must be that
star formation without metals produces less massive stars then
the present processes with metals.  It is tempting to make a simple
argument that the power law spectra that allow many massive
stars to be produced are consequences of self-organized critical
phenomena that require a certain chemistry, and hence complexity.
But it is also clearly a possibility that such an argument would be too
naive.

Let me describe one more prediction made by this theory.  One
parameter that plays a crucial role in determining the number of
black holes produced is the upper mass limit for neutron
stars, $m_{uml}$. A supernova remnant becomes a black hole if it
is more massive than this, otherwise it becomes a neutron star.  The
theory I described must predict that this parameter is as low as
possible, consistent with other processes that play a role in
star formation.  What would be especially interesting is if
$m_{uml}$ were under the control of a parameter that played a
minimal
role in the star formation processes or early universe cosmology, for
if this were the case, it could be independently varied and
minimized,
to maximize the number of supernova remnants that become black
holes.

Remarkably, it seems that there is such a parameter: it is the
strange quark mass.  The reason is that, according to calculations
of Brown, Bethe and their collaborators\cite{bethebrown},
if the mass of the kaon is
low enough, the neutron star matter will be dominated by a kaon
condensate.  This turns out to greatly soften the equation of state
from what it would be if the condensate were absent, which in turn
lowers $m_{uml}$.
The result is that they predict $m_{uml}=1.5 M_{solar}$, while
conventional equations of state lead to $m_{uml}=2.5-3 M_{solar}$.

If their general arguments are correct, then there is a value
of $m_{strange}$, the strange quark mass, $m_{critical}$,
such that for $m_{strange} < m_{critical}$ the condensate
dominates neutron stars.  The question is whether the actual
$m_{critical}$ is above the actual value of $m_{strange}$.
I may note that the theory I've described here predicts that
it must be, for if nature had the possibility of choosing
$m_{strange}$ so that many more black holes were produced, and
didn't use it, the theory is definitely wrong.

Thus, on this theory I must predict that in fact
$m_{uml}=1.5 M_{solar}$.  Thus, the discovery of one neutron star
with a larger mass would be strong evidence against it.
In fact, of about 18 neutron star masses that are so far
measured,
all are within error below this value\cite{nsmasses}.

But there is a second question, why is $m_{uml}$ not still lower?
If it were, many neutron stars would instead be black holes.
If the theory is true then there must be competing effects that
prevent $m_{uml}$ from being lowered still further, even if
$m_{strange}$ is lowered.  This is a question that can be
investigated theoretically, and work on it is underway.

While this discussion has been sketchy, I hope to have convinced
the reader that the idea that quantum gravity has no experimental
consequences is a bad rap.  Here we find that two very plausible
assumptions about what happens inside of black holes at the Planck
scale result in predictions that can be tested by both observational
and theoretical work in astronomy and nuclear physics.

It is quite possible, perhaps even likely that this particular
theory is wrong, as I've emphasized if it is wrong we will
be able to tell.  But we may still learn something from it,
for this coupling of assumptions about the Planck scale to
predictions about things we can observe is exactly what we
may expect if we go away from the idea that the parameters of
physics and cosmology are picked by some mathematical principles
acting at the Planck scale, and move in the direction of a theory
in which they are determined by real mechanisms of self-
organization
that may have occurred some time in our past.

\section{Critical phenomena in quantum gravity and the classical
world}

Now I would like to come to another way in which critical
phenomena
are likely to play an important role in cosmology.   This application
is different from the others I've described, as it involves directly
the physics of the Planck scale.  As I mentioned earlier, if one
assumes that the universe expanded from an initial state, with
temperature and densities given by natural units in particle
physics, it becomes difficult to understand
how the universe managed to expand to the present size, without
either collapsing or entering a phase of runaway expansion.
However, as I will describe here, the actual situation may be
even worse than this.  Recent developments in quantum gravity
suggest that even the fact that the world has scales in it
significantly larger
than the Planck scale, which is
necessary if it is to be  describable in terms
of classical geometry, is improbable without
fine tunings\footnote{It may be emphasized that in quantum
gravity the classical limit is the same as a limit of large
distances because $\hbar$ appears only in the Planck
length, $l_P=\sqrt{\hbar G_{Newton}}$.  Equivalently, it
makes no sense to speak of a classical
description at the Planck
scale.}.  Just
the fact that there is a world describable in terms of
classical space and time, I will argue, is a problem in
critical phenomena.

Let me first make the one paragraph
argument that this might be the
case, then I will show that this argument does in fact correspond
to what we know about quantum gravity.
A quantum theory of gravity has one scale in it, the Planck
scale.  Because the scale is also the gravitational coupling
constant, what a quantum theory of gravity naturally describes
is a strongly coupled phase in which there are no correlations
on larger scales.  But as a quantum theory of gravity is
a theory of geometry, the existence of a semiclassical limit
means that there is a description in terms of a classical geometry
in which the averages of classical curvatures are small in
Planck units.  Thus, classical space and time are themselves
consequences of a critical behavior in which there are correlations
on scales much larger than the Planck scale.  Further, as
the coupling
of excitations of the geometry are proportional to the wavelength,
in Planck units, the existence of a classical limit in a quantum
theory of gravity means precisely that the system is critical
and weakly coupled.  Generically, such a phase cannot exist
naturally unless there is some reason for the system to be critical.

Perhaps one might have the impression that this argument
proves too
much.  For what it claims is that in any formulation of quantum
gravity in which the existence of classical spacetime is not put
in from the beginning, it will be hard to get the classical world
out, unless the theory has a critical point for some tuning of the
parameters or initial conditions.  Formulations of quantum
gravity that do not assume that the world is described by
small perturbations around a classical spacetime are non-
perturbative
by definition.   And, so far, every non-perturbative formulation that
has been developed sufficiently to ask the question leads to
the picture I've described.

This has been seen in both path integral and hamiltonian
formulations
of non-perturbative quantum gravity.  In the path integral case,
non-perturbative calculations have been performed by discretizing
the manifold, and then averaging over
a certain set of discrete geometries,
as in the case of random surface models in lower
dimension\cite{review-triangles}. There are two such
formulations, the dynamical triangulation
models, developed by Agishtein and Migdal\cite{AM} and
Ambjorn and collaborators\cite{Ambjorn}
that mimic closely the random surface theory and the
Regge calculus models\cite{reggemodels},
which use an older approach in which the dynamical variables are
the
edge lengths of a fixed triangulations\cite{regge}.

The results are similar in these two cases.  The models have two
parameters, which correspond to Newton's constant, $G$,
and the cosmological
constant, $\Lambda$.
There are two phases, a crumpled phase in which macroscopic
distances are not defined, and the Haussdorf dimension grows with
the
size of the system, and an elongated phase, in which things are
greatly stretched out, so that the Hausdorff dimension of spacetime
is close to $2$.  Between them there is a second order phase
transition governed by a non-trivial critical point at which
the Haussdorf dimension seems to be four, within statistical error.

So in these models the picture I described is exactly true.
Despite the fact that it is constructed by making a discrete
approximation to four dimensional general relativity, the theory
can only predicts the existence of a classical four dimensional
spacetime when the parameters are tuned to a
critical point\footnote{The general idea that the existence
of four dimensional quantum gravity would require the
presence of a non-trivial scaling behavior associated
with a non-Gaussian fixed point was anticipated some
time ago on general grounds\cite{fixed}}.

A similar picture emerges from the Hamiltonian formulation.
Without going into details, one approach to the Hamiltonian
quantization of general
relativity\cite{cr-review,aa-review,ls-review,weaves,volume}
has advanced to the point
that the following simple picture has emerged:

The quantum states of the gravitational field are in one
to one correspondence with a certain class of graphs, which
are called spin networks\cite{roger}.
These are graphs in which the edges
are labeled by half-integers corresponding to spin, and the
laws of addition of angular momentum must be satisfied at vertices.
It should be emphasized that the graphs are defined only
topologically, they are not located anywhere in space, because
they are the quantum fields that comprise space.

These states have a simple physical
interpretation\cite{volume,spinnet-us}:
they are eigenstates
of certain observers that measure the geometry of space by
determining
the areas of arbitrary surfaces and the volumes
of arbitrary regions.  Given any such graph, one may draw
regions and surfaces and assign them areas and volumes according
to simple rules.  Every surface has an area given by the sum of
the spins on the edges of the graphs that intersect it, in units
of the Planck length squared.  Every vertex carries a certain
discrete amount of volume, given by a certain combinatorial
formula of the spins that enter it, times the Planck length cubed.

I want to emphasize that this simple picture was not dreamed up,
it is the result of calculations.  The fact that the operators that
measure
physical areas and volumes are discrete is a prediction of quantum
general relativity.

Given such a network then, there is a discrete geometry, which is
somewhat analogous to those that are integrated over in the
path integral approaches (only they correspond to space and not
spacetime.)  As in that case, almost none of the states of the
theory correspond to smooth classical geometries.  For certain
very special states, based on very large networks which satisfy
certain conditions of regularity, it is possible to describe
the geometry on the average in terms of a classical metric.
But the conditions that make this possible are rather strict, and
most of the states of the system do not correspond to any
classical geometry, nor do they define any scale of phenomena
larger than the Planck scale.

The dynamics under which these networks evolve has been worked
out,
given certain assumptions about time.  This is a long story in
itself, let me say only that time here is measured relative to some
matter field\cite{ls-dieter}.
The hamiltonian is known, and is a finite, well
defined operator\cite{ham1}.  Its action is particularly
simple when developed in a strong coupling expansion, in a
dimensionless parameter which is $1/G^2 \Lambda$.  There
are processes that turn
vertices into little triangles by adding two new vertices, and
processes that do the reverse and
collapse little triangles to nodes\cite{ham2}.

The description is very beautiful, and calculations of transition
amplitudes can be carried out to any order in this strong coupling
expansion.  The problem, of course, is that the dynamics in
this strong coupling phase does not seem to correspond to the
weak coupling picture in which massless gravitons move on a
background described by a classical spacetime.

I should emphasize that the problem is not with the existence
of gravitons {\it per se.}  It is known, in fact, that if one
can assume the existence of a state that has a classical
description in terms of a flat geometry, its long wavelength
excitations consistent with the gauge invariance and dynamics
are precisely two massless spin two modes per
momenta\cite{weavegravitons}. The
problem is that the theory does not naturally predict the
existence of a state associated with a classical geometry.

I might stress that this is an intrinsically cosmological problem,
in that a boundary condition has been imposed in which the universe
is spatially compact.  This was a condition that Einstein
argued for on philosophical grounds, as he invented the science
of relativistic cosmology.  He was motivated to do so by
the philosophical tradition of Leibniz and Mach according
to which space and time should not exist a priori, but should
be a consequence of dynamical relations among things in the
world.  What seems to be the case is that
when quantum theory is added to the picture this philosophy is
realized precisely in that all that one has for generic
couplings is a description of a dynamically evolving network
of relations.  That these have long range correlations such that
space and time exist at all has become a dynamical problem, it has
become precisely a problem of critical phenomena.

As I said in the introduction, we understand two broad classes of
critical phenomena, second order phase transitions and self-
organized
critical phenomena.  The first requires that  parameters be tuned
to a critical point. But we are discussing a theory that is supposed
to be a fundamental theory of cosmology.  We might then argue that
in such a theory it is not acceptable to explain the existence of the
classical world by means of a delicate tuning of parameters.  There
is nothing outside the world that can tune the parameters.  Thus,
if it is to succeed, quantum cosmology must become a study of a
self-organized critical phenomena.  There must be a natural
mechanism of self-organization that explains why the quantum
state of the world is in an improbable critical state.

Perhaps this may seem too philosophical.  But we must keep in my
mind that any such theory may be observationally testable, for we
may expect generally that if there is a mechanism of self-
organization
that explains naturally why the world gets big and classical, that
mechanism is likely going to produce a scale invariant spectrum
of fluctuations around the average state.  Thus, such a mechanism
is likely to produce an outcome similar to that given by
inflationary cosmologies, which is a large classical world on which
there is an approximately scale invariant distribution of fluctuations,
but, if it succeeds, it will do it naturally, without the fine tunings
required to make inflationary models work.  As such, it is likely
to make testable predictions about the details of the fluctuations
seen in
the microwave background radiation.

\section{Variety, complexity and relativity}

It is of course possible that the point of view I've sketched
in the last sections will not turn out to be useful.  The test
of any scientific hypotheses must,
in the end, can be nothing other then whether they work out
in detail to explain the empirical world.  Thanks to the work of
the astronomers, cosmology is becoming more and more a
question of the details.  But, even so, I would like to argue that
what is happening deserves some wider reflection.  I offer
the following as a possible point of view, for whatever it may
turn out in the end to be worth.

What we are engaged in is an attempt to make sense of a
cosmological theory based on general relativity and quantum
theory.   This, I would like to argue, must lead to a description
of a world that is intrinsically complex, so that the complexity of
the world we see must be not accidental, not a matter of a fine
tuning of parameters, but in some way inherent in the postulates
of quantum theory and relativity.

I know of two arguments for this, one from
relativity and one from quantum theory.

The argument that the principles of relativity require a complex
world, when applied in a cosmological context  is based
only on diffeomorphism invariance, which is the most fundamental
principle of general relativity.  It is the gauge symmetry
of the theory,
thus it has a more secure status then the particular forms
of the dynamical equations.  We might expect
that it could be included in a
larger gauge symmetry in some unification such as the posited
non-perturbative string theory, but we cannot expect
general relativity to
be unified into a more fundamental theory without
diffeomorphism invariance.

Diffeomorphism invariance, which Einstein called general
covariance, has a very simple meaning in the context of field
theory.  It says that points
have no meaning unless they are described by the values
of physical fields.    No physical observable can speak about
what happens at a point of spacetime, unless that point is
determined uniquely by the fields that an observer at that
point would measure.  You cannot say, what is the curvature
scalar at point $x$.  You can only say something like:
what the value of the
scalar curvature is at  a point
where the value of the electromagnetic fields (and perhaps their
derivatives) are such and such\footnote{There has
been in the past
some controversy about the question
of the interpretation of general relativity, but this view
is presently widely understood by relativists to be
correct.  That it was
Einstein's point of view is convincingly shown by
Stachel in \cite{stachel-hole}.  This point of view has also
been found to be necessary to make progress in quantum
gravity\cite{cr-review,ls-review,ls-dieter,volume}.}.

Like any gauge theory, the physical interpretation of
general relativity must be described in terms of gauge invariant
observables. As the theory has two degrees of freedom per point,
there must be an equal number of such observables.  They must
all be complicated functions that describe relationships between
fields, such as I have described.

Now we come to the key point, which is that
such observables will not be well defined for a given
cosmological solution to
the theory unless it describes a world that is complex enough
that points of spacetime can be uniquely described by the
values of the fields there.  This has a simple
consequence, it means that to have a good,
gauge invariant interpretation, a spacetime must be complex
enough that no two observers observe exactly the same thing,
no matter where they are in space and time.  To put it more
informally, it must be possible to tell where in the world you
are, and when it is, uniquely from what you see when you look
around you.

We live in a world with enough variety and structure that this
is certainly the case.  What I am arguing is that if the gauge
invariance of the world includes diffeomorphism invariance
this cannot be an accident: it is required if the theory is to
have a good interpretation.

There may seem to be a problem with this argument, which is
that no solution with symmetry can be given a good physical
interpretation by means of such observables, precisely because
a symmetry means that there are points that are not distinguished
by the values of the fields.   But we use solutions with symmetries
all the time to model relativistic cosmologies, and we are able
to interpret them.  Certainly we are, but we do this in a way that
makes use of special coordinate systems that are present because
of the symmetries.  These methods do not generalize to other
solutions, nor, I am claiming, can any interpretation that applies
generally to relativistic cosmologies be applied to the symmetric
solutions.

What we are really doing when we study solutions with symmetries,
of course, is taking advantage of the fact that the symmetry is not
exact, for it is only by the detailed distribution of matter,
that break it, that we are able to give
meaning to the coordinates we use.

This circumstance would not be a problem in a Newtonian cosmology,
as coordinates are intrinsically meaningful according to the
Newtonian conception of space and time.  But general relativity is
in a different tradition, it is in the tradition of Leibniz and Mach,
who argued for a view of space and time in which they are only
meaningful to the extent that they are seen in relationships
between real things.  Indeed, Leibniz understood from the
beginning that any cosmological theory in which such a view
of space and time was realized would have to describe a world
with sufficient complexity that no two observers have exactly the
same view of things\cite{monadology}.

The second argument for a complex world, coming from quantum
theory has been given by many others, so I will be brief.
Quantum theory does not seem to make sense unless there are
observers in the world.  Therefor, any quantum theory that
successfully applies to cosmology must, by self-consistency alone,
describe a world complex enough to have observers.

In my opinion, the first argument is stronger than the second.
It could easily turn out that quantum theory cannot be extended
from the microscopic world to the cosmological.  But the first
argument uses the most secure principle of general relativity which
is diffeomorphism invariance.  The observed orbits of the binary
pulsars show that we live in a world in which the geometry of
spacetime is dynamical, which means there can be no going back
to the Newtonian conception of space and time.

However, given either argument we reach the conclusion that
a cosmology which is consistent with both general relativity and
quantum theory must, by self-consistency alone, describe a
complex universe.

If this is, however, to be a good scientific argument, it must be
possible to make it quantitative.  There ought to be a measure
of the complexity of the universe, or of any closed system, that
describes how easily each observer may be distinguished by
their view of the rest of the system.  I would then like to close
by describing one
such approach to a quantitative measure of complexity, that Julian
Barbour and I have been developing.

To define such a notion, we need a system, made of a number
of elements, which I will denote $x_i$.  One can think of
these as particles or observers, as one likes.  What is required
is that there be a space $\cal w$ that
contains the  possible views of the
system.  To each element $x_i$
we are able to construct an element, $v_i$ that can be
called its view of the system.

For example, the system could be a lattice dynamical system
in $D$ dimensions, in which case an element of $\cal V$
consists of a series of spaces ${\cal V}_n$ which
describe the possible configurations of
neighborhoods of a point in the lattice.
$n$ refers to the number of steps away from the original
point that describe the neighborhood so that
${\cal V}_n$ is the space of possible configurations of
a $(2n+1)^D$ lattice of points $n$ steps away from a given
site.

Another possibility is that the system is a graph
or a network,
perhaps of the kind we discussed in the previous section, in
which case the neighborhoods  ${\cal V}_n$ are all the subnetworks
with a distinguished point, corresponding to the element,  which
contain points up to $n$ steps away from it.

Still another possibility is that the system consists of
$N$ points  distributed in $D$ dimensional
space, in which case its view of the rest of the system are
$N-1$ points distributed on a $D-1$ dimensional sphere that
describes where it sees the other points on its sky.

 Given any such system, which defines a set of views
$w_i$ of each element, we may define the {\it variety}
of the system as follows.

We must first construct a matrix of differences
$D_{ij}$ that measure how far apart the views of
the $i$'th and $j$'th elements are from each other.
There are two approaches to this.  The space of
views could be a vector space, in which case
\f
D_{ij} = |w_i -w_j |
\ff
Or, in the cases in which the views comprise a sequence of
neighborhoods, the difference $D_{ij}$ is simply
$1/n_{ij}$, where $n_{ij}$ is the smallest
$n$ such that the two $n$ step neighborhoods are different.

Given the matrix of differences, the variety of the
system may be defined.
\f
V= \sum_{i \neq j } D_{ij}
\ff

We have applied this definition of complexity to a number of
systems, including graphs and points in one and two dimensional
spaces\cite{variety}.  We find that systems
that have high variety are
generically distinguished by being complex without being
ordered, so that any two points can indeed by easily
distinguished from each other by looking at what is around
them.  Moreover, this is a definition of complexity that
distinguishes true complexity from order, for ordered
configurations, or configurations with any kind of symmetry
turn out to have low variety.  Generically, we find that
ordered configurations have much lower varieties than randomly
generated configurations, while configurations with high variety
are easily distinguished from both ordered and random
configurations.

Thus, the variety of a system may be defined quantitatively.
The next step is to try to define an appropriate notion
of variety for classical or quantum general relativity.  We may,
for example,
try to define the variety of a quantum spacetime to be
inversely proportional to the average number of
bits of information an observer
must have in order to locate themselves
 uniquely in space and time.
We may then conjecture that the dynamics of a quantum
gravitational theory act to increase the variety of
typical configurations in time.  Certainly,
as gravitation acts to
form hierarchies of bound systems, as we see from the
Press-Schecter model, and more generally makes it
possible for large regions of
the world to be kept far from thermal equilibrium for
arbitrarily long periods, this is not obviously wrong.  If true,
this would be a step towards a picture in which we understood
that our world is organized because a quantum gravitational
system must, for its own self-consistency,
contains intrinsic statistical mechanisms of self-organization.

\section*{ACKNOWLEDGEMENTS}

These notes are based on a series of lectures I was asked to
give at the Guanajuato workshop ``Complex systems and binary
networks."
I want to thank the organizers of the Guanajuato
school for giving me the opportunity to describe these thoughts in
front of a knowledgeable and critical audience.  I want to also
thank the audience and lecturers  for many
illuminating
discussions and criticisms.  Some of
the ideas described here have been
developed over a long period of time
through conversations and collaborations with Julian
Barbour, Jane Charlton, Louis Crane and Carlo Rovelli.  I would
also like to thank also Per Bak, Gerry Brown, Saint Clair
Cemin, Freeman
Dyson, James Peebles,
Martin Rees, Stanley Rosen, Larry Schulman and John
Wheeler for helpful
discussions about these problems.
I am also grateful to Jane Charlton for a critical reading
of the manuscript.
Finally,  I would like to thank
the astrophysicists of the Institute for Advanced Study, where
these notes were written,
for an atmosphere most conducive to
reflection on the unsolved problems in cosmology.
This work
has been supported in part by the NSF grant  PHY-93-96246.


\begin{thebibliography}{99}

\bibitem{Omega-WNE}S. D. M. White, J. F. Navarro, A. E. Evrard,
C. S. Frenk, Nature 366 (1993) 429-433;

\bibitem{ColesEllis}P. Coles and G. Ellis {\it The case for an
open universe} Department of Applied Math prprint, Capetown
(1994).

\bibitem{numass}See for example, J. Bachall, in the Proceedings of
Some Unsolved Problems in Astrophysics.

\bibitem{COBE} Mather, J. C. et al. Ap. J. 354 (1990) L37.

\bibitem{structure} V. De Lapparent,  M.J. Geller and J.P.
Huchra Ap.J. 302 (1986) L1;  M. P. Haynes and R. Giovanelli,
Ap. J. 306 (1986) L55.

\bibitem{garay} L J For an excellent review, see
Garay: ``Quantum gravity and minimum
length'', Imperial College preprint/TP/93-94/20, gr-qc/9403008
(1994)

\bibitem{Hoyle}F. Hoyle, D. N. F. Dunbar, W. A. Wensel and
W. Whaling, Phys. Rev. 92 (1953) 649; F. Hoyl, {\it Galaxies,
Nuclei and quasars} (Heinemann, London, 1965), p. 146.

\bibitem{CarrRees}B. J. Carr and M. J. Rees, Nature 278 (1979) 605.

\bibitem{BarrowTipler}J. D. Barrow and F. J. Tipler
{\it The Anthropic Cosmological Principle}
(Oxford University Press,Oxford,1986).

\bibitem{Carter}B. Carter, {\it "The significance of
numerical coincidences in nature"}, unpublished preprint,
Cambridge University, 1967; in {\it Confrontation of
Cosmological Theories with Observational Data},
IAU Symposium No. 63, ed. M. Longair (Reidel, Dordrecht,1974)
p. 291.

\bibitem{BTW}P. Bak, C. Tang and K. Wiesenfeld, Phys. Rev. A38
(1988) 364; Phys. Rev. Lett. 59 (1987) 381.

\bibitem{per-soc}P. Bak and Maya Paczuski, ``Complexity,
contingency and criticality" Brookhaven preprint.

\bibitem{soc-accretion}S. Mineshinge, N. B. Ouchi and H. Nishimori,
PASJ 46 (1994) 97;
S. Mineshinge, M. Takeuchi and H. Nishimori,
Ap. J. 435 (1994) L125.

\bibitem{IBMguys}H. Gerola and P. E. Seiden,
Ap. J. 223 (1978) 129; P. E. Seiden, L. S. Schulman and H. Gerola,
{\it Stochastic star formation and the evolution of galaxies},
Astrophys. J. 232 (1979) 702-706; P. E. Seiden
and L. S. Schulman, {\it Percolation and galaxies}
Science 233 (1986) 425-431 {\it Percolation model of galactic
structure},
Advances in Physics, 39 (1990) 1-54; L. S. Schulman,
``Modeling galaxies:: cellular automana and percolation",
to appear in {\it Cellular Automata: Prospects in Astrophysical
Applications}, A. Lejeune and J. Perdang, eds. World Scientific,
Singapore (1993).

\bibitem{ssg-morphology}P. Seiden, L.S. Schulman and H. Gerola,
Ap. J. 232 (1979) 702.

\bibitem{spiralstructure}See, for example:
J. Franco and D. P. Cox, {\it Self-regulated star formation in the
galaxy}  Astrophys. J. 273 (1983) 243-248;  J. Franco and S. N. Shore
{\it The galaxy
as a self-regulated
star forming system:  The case of the OB associations} Astrophys. J.
285 (1984) 813-817;
S. Ikeuchi, A. Habe and Y. D. Tanaka {\it The interstellar medium
regulated by supernova remnants and bursts of star formation}
MNRAS 207 (1984) 909-927;
R.F.G. Wyse and J. Silk {\it Evidence for supernova regulation of
metal inrichment in disk galaxies} Astrophys. J. 296 (1985) l1-l5;  M.
A. Dopita,
{\it A law of star formation in disk galaxies: Evidence for self-
regulating
feedback} Astrophys. J. 295 (1985) L5-L8;
G. Hensler and A. Burkert, {\it Self-regulated star formation and
evolution of the
interstellar medium} Astrophys. and Space Sciences 171 (1990) 149-
156.

\bibitem{WyseSilk}R. F. G. Wyse and J. Silk, Astrophys. J.
339 (1989)
700.

\bibitem{clouds}See, for example, {\it The Physics and
Chemistry of
Interstellar
Molecular Clouds}  ed. G. Winnewisser and J.T. Armstrong, Springer
Verlag
Lecture Notes in Physics 331(1989);  {\it Molecular Coulds in the
Milky Way
and External Galaxies}  ed. R. L. Dickman, R. L. Snell and
J. S. Young
Springer Verlag Lecture Notes in Physics 315 (1988).

\bibitem{elmegreen-triggered}B. G. Elmegreen {\it Triggered Star
Formation}  IBM Research Report, in the Proceedigs
of the III Canary Islands Winter School, 1991,
eds. G. Tenorio-Tagle, M. Prieto and F. Sanchez (Cambridge
University Press, Cambridge, 1992).

\bibitem{elmegreen-review}B. G. Elmegreen, {\it Large Scale
Dynamics of
the Interstellar Medium}, to appear
in {\it Interstellar Medium, processes in the galactic diffuse
matter}
ed. D. Pfenniger and P. Bartholdi, Springer Verlag, 1992.

\bibitem{parravano}A. Parravano, {\it Self-regulating star
formation
in isolated galaxies: thermal
instabilities in the interstellar medium}  Astron. Astrophys. 205
(1988) 71-76;  {\it A self-regulated star formation rate as
a function
of global galactic parameters}  Astrophys. J. 347 (1989) 812-816;
A. Parravano and J. Mantilla Ch., {\it A self-regulated
state for the
interstellar medium: radial dependence in the galactic plane},
Atrophys. J. 250 (1991) 70-83;
A. Parravano, P. Rosenzweig and M. Teran, {\it Galactic evolution
with
self-regulated star formation: stability of a simple
one-zone model}
Astrophys. J.
356 (1990) 100-109.

\bibitem{stars}See, for example, F. H. Shu, F. C. Adams and S.
Lizano
{\it Star formation in molecular clouds: observation and theory}
in
Ann.
Rev. Astron. Astrophy.  25 (1987)  23-81 and C. J. Lada and F.
H. Shu,
{\it The formation
of sunlike stars} Berkely preprint, to appear in Science and
references contained
therein.

\bibitem{Salpeter}E. E. Salpeter, Astrophys. J. 121 (1955) 161.

\bibitem{Scalo}G. E. Miller and J. Scalo, Ap. J. Suppl. 41
(1979) 513;
J. Scalo, Fundamentals of Cosmic Physics 11 (1986) 1-278.

\bibitem{Larson} R. B. Larson M.N.R.A.S.  214 (1985) 379;
218 (1986) 409.

\bibitem{scalo-fractal}J. Scalo, in {\it Physical Processes in
Fragmentation and Star Formation} ed. R. Capuzzo-Dolcetta, C.Ciosi
and A. Di Fazio (Klower,1990)

\bibitem{elmegreen-model}B. Elmegreen and M. Thomasson,
{\it Grand design and flocculent spiral structure in computer
simulations with star formation and gas heating} Astron.
and Astrophys. (1992) ?.

\bibitem{openinflation}M. Bucher, A. S. Goldhaber and N.
Turok, ``An open universe from inflation",
hep-ph/9411206, iassns-hep-94-81. PUPT-94-1507;
M. Bucher and N. Turok, ``Open inflation with arbitrary false
vacuum mass" hep-ph 9503393, PUPT-95-1518;
J.R. Gott, Nature 295 (1982) 304.

\bibitem{Walker}Walker, T. P, Steigman, G., Kang, H.-S.,
Schramm, D. M., \& Olive, K. 1991,
%``Primordial Nucleosynthesis Redux'', ApJ, 376, 51

\bibitem{models-lss}See, for
example, S. D.M. White, MAP preprint, 1994;
S. D. M. White and C. S. Frenk, Ap. J. (1991) 379, 52;
G. Efstathios and J. Silk, Fund. Cos. Phys. 9 (1983) 1.

\bibitem{oc}J.P. Ostriker and L.L. Cowie, Ap.J. 243 (1981) L127.

\bibitem{ikeuchi}S. Ikeuchi, Publ. Astron. Soc. Japan
33 (1981) 211.

\bibitem{osetal}J.P. Ostriker, C. Thompson and E. Witten,
Phys. Lett. B (1986).

\bibitem{daley}R.A. Daley (1986)

\bibitem{OstGen}N. Yu. Gnedin and J. P. Ostriker,
Astrophys. J. 400 (1992) 1-20

\bibitem{Peeblesbaryonic}P.J.E. Peebles, in
{\it The Early Universe} ed. W.G. Unruh and G.W. Semenoff,
D. Reidal Publishing, 1988, p. 203; in the proceedings
of the 8th IAP meeting, {\it First light in the universe}

\bibitem{hogan}C. Hogan, Ap. J. 415 (1993) L63-66.

\bibitem{qsoabsorb}P. Petitjean, J. K. Webb, M. Rauch, R.F.
Carswell and K. Lanzetta, MNRAS 262 (1993) 499;
K.M. Lanzetta, A. M. Wolfe, D. A> Turnshek, Limin Lu,
R.G. McMahon and C. Hazard, Ap. J Suppl. Series.
77 (1991) 1.

\bibitem{jane-break}J. Charlton, E. Salpeter and C. J. Hogan,
Ap. J.
402 (1993) 493; J. Charlton, E. Salpeter and S. M. Linder,
``Competition between pressure and gravity confinement in
Lyman alpha forest observations", ApJ, 430, L29 (1994).

\bibitem{qsocorrelations}Lanzetta, K. M., Bowen,
D. V., Tytler, D., and Webb, J. K.
1995, ApJ, 442, 538; Steidel, C. 1995, in
Proceedings of ESO Workshop on QSO Absorption Lines,
ed. G. Meylan, (Springer-Verlag: Heidelberg), in press

\bibitem{noevolution}Richard Ellis ``The morphological
evolution of galaxies" in {\it Unsolved Problems in
Astrophysics} {\it op. cit.}

\bibitem{dwarfs}Broadhurst, R. J., Ellis, R. S., \& Shanks,
T. 1988, MNRAS, 235, 927
Colless, M. M., Ellis, R. S., Taylor, K., \& Hook, R. N.
1989, MNRAS, 244,408
Songaila, A., Cowie, L. L., Hu, E. M., \& Gardner, J. P.
1994, ApJS, 44, 461

\bibitem{peebles-book}J. Peebles, {\it Principles of Physical
Cosmology} (Princeton University Press,1993).

\bibitem{structuremodels}For a general review, see
N.Bachall and J. Ostriker, in the Proceedings of the Conference
on Unsolved Problems in Astrophysics, {\it op. cit.}.
See also R.Y. Cen and J. P. Ostriker, Ap. J.
339 (1992) L113; 404 (1993) 415; 417 (1993) 415;
D. Ryu, J.P. Ostriker, H. Kang and R.Y. Cen, Ap.J. 414 (1993) 1

\bibitem{MauroKamilla}M. Carfora and K.
Piotrokowska, ``A renormalization group approach to
relativistic cosmology", to appear in Phys. Rev. D.

\bibitem{pressschecter}W. H. Press and P. Schecter,
Ap. J.. 187 (1974) 425.

\bibitem{support-ps}S. D. M. White, G. Efstathiou and C. S.
Frenk, Mon. Not. R. Astro. Soc. 262 (1993) 1023;
A. Klypin, J. Holtzmann, J. Primack and E. Regos Ap. J. 416
(1993) 1;
Lacey and Cole, MNRAS (1994)

\bibitem{ChenBak}K. Chen and P. Bak, Phys. Lett. A 140 (1989) 299.

\bibitem{SchulmanSeiden}L. S. Schulman and P. E. Seiden,
Ap. J. 311 (1986) 1.

\bibitem{carr-dark}B.J. Carr, ``Baryonic dark matter",
to appear in {\it Annual Reviews of Astronomy and
Astrophysics}, 1995.

\bibitem{andy}A. Strominger, ``Massless black holes and
conifolds in string theory", preprint hep-th/9504090.

\bibitem{evolution1}L. Smolin Classical and Quantum Gravity 9
(1992) 173-191

\bibitem{evolution2}L. Smolin, {\it On the fate of black hole
singularities and the parameters
of the standard model} gr-qc ??

\bibitem{book}L. Smolin, {\it The Life of the Cosmos} to
appear in Oct. 95, Crown Press,New York, and Orion Press, London.

\bibitem{wheeler}J.A. Wheeler, in {\it Gravitation},
by C. Misner, K. Thorne and J. A Wheeler, last chapter.

\bibitem{martinec}E. Martinec, 1994, hep-th/9412074

\bibitem{gott-talk}R. Gott, private communication.

\bibitem{bethebrown}G. E. Brown and H. A. Bethe, Astro. J. 423
(1994) 659; 436 (1994) 843, G. E. Brown, Nucl. Phys. A574
(1994) 217; G. E. Brown, ``Kaon condensation in dense matter";
H. A. Bethe and G. E. Brown, ``Observational constraints on
the maximum neutron star mass", preprints.

\bibitem{nsmasses}S. E. Thorsett, Z. Arzoumanian, M.M. McKinnon
and J. H. Taylor  Astrophys. Journal Letters 405 (1993) L29

\bibitem{rees-fragment}M. Rees, MNRAS 176 (1976) 483;
J. Silk Ap.J. 211 (1976) 638.

\bibitem{review-triangles}For a review, see
J. Ambjorn, J. Jerkiewicz and
Y. Watabiki, ``Dynamical triangulations, a gateway to quantum
gravity" NBI-HE-95-08, to appear in J. Math. Phys. Nov. 1995.

\bibitem{AM}M.E. Agishtein and A. A. Migdal, Nucl. Phys. B385
(1982) 395.

\bibitem{Ambjorn}J. Ambjorn, J. Jerkiewicz and C. F. Kristjansen,
Nucl. Phys. B393 (1993) 601; Phys.  Lett. B305 (1993) 208;
J. Ambjorn, Z. Burda, J. Jerkiewicz and C. F. Kristjansen,
Phys. Rev. d48 (1993) 3695.

\bibitem{reggemodels}H. W. Hamber, Nucl Phys. B (Proc. Supp.)
20 (1991) 728; 25A (1992) 150; B400 (1993) 347;
Phys. Rev. D45 (1992) 507; H. W. Hamber and R. M. williams,
Nucl. Phys. B415 (1994) 463.

\bibitem{regge}T. Regge, Nuovo Cimento 19 (1961) 558.

\bibitem{fixed}S. Weinberg, in {\it  General Relativity:
An Einstein Survey}
ed. S. Hawking and W. Israel (Cambridge University Press,1979).
L. Smolin,   Nuclear Physics B208  (1982) 439.

\bibitem{cr-review}  C Rovelli: Class Quant Grav 8 (1991) 1613

\bibitem{aa-review} A  Ashtekar:  {\it Non perturbative canonical
gravity}, World scientific, Singapore 1991

\bibitem{ls-review}L  Smolin: in {\it Quantum Gravity and
Cosmology}, eds  J  P\'erez-Mercader {\it et al}, World Scientific,
Singapore 1992

\bibitem{weaves}  A Ashtekar C Rovelli L Smolin: Phys Rev Lett
69 (1992) 237

\bibitem{volume}C. Rovelli and L. Smolin, Discreteness of volume
and
area in quantum gravity, to appear in Nucl. Phys. B 1995.

\bibitem{roger}R Penrose: in {\it Quantum theory and
beyond}  ed T Bastin, Cambridge U Press 1971;
in {\it Advances in Twistor Theory}, ed. L. P. Hughston and R. S.
Ward,
(Pitman,1979) p. 301; in {\it Combinatorial Mathematics and
its Application} (ed. D. J. A. Welsh) (Academic Press,1971).

\bibitem{spinnet-us}C. Rovelli and L. Smolin,
``Spin networks and quantum gravity" Penn State
 CGPG-95/4-4 and IASSNS-HEP-95/27 preprint,
gr-qc/9505006.

\bibitem{ls-dieter}L Smolin: in {\it Directions in
General Relativity,
v. 2, papers in honour of Dieter Brill}, ed BL Hu T Jacobson,
Cambridge  University Press, Cambridge 1994


\bibitem{ham1}C. Rovelli and L. Smolin,
Phys Rev Lett 72 (1994) 446

\bibitem{ham2}R. Borissov, C. Rovelli and L. Smolin,
{\it Nonperturbative dynamics of quantum general relativity}
preprint in preparation.  C. Rovelli, to appear in J. Math
Phys. Nov. (1995).

\bibitem{weavegravitons}J Iwasaki
C Rovelli: Int J of Mod Phys D1 (1993) 533;
Class and Quantum Grav 11 (1994) 1653

\bibitem{stachel-hole}J. Stachel, {\it Einstein's search for
general covariance 1912-1915}  in {\it
Einstein and the History of General Relavity} ed. by D.
Howard and J.
Stachel,
Einstein Studies, Volume 1  (Birkhauser,Boston,1989).

\bibitem{monadology}Leibniz, {\it The Monadology} in
{\it Leibniz,
Philosophical Writings}  ed.
G.H.R. Parkinson, translated by M. Morris and G.H.R. Parkinson
(Dent,London,1973)

\bibitem{variety}
J. B. Barbour and L. Smolin, Syracuse University preprint,
SU-GP-92/2-4,
see also
J. B. Barbour, "On the origin of structure in
the universe,"  presented at the "3d Philosophy and Physics
Workshop", Forschungsstatte der Evangelischen
Studiengemeinschaft (FEST) in Heidelberg, May 1990.  To
be published in "Philosophy and Modern Physics",
publ. by Springer; {\it Mathematical
Modeling of the Monodology}, submitted for publication
and L. Smolin {\it Space and Time in the Quantum
Universe}  in {\it Conceptual Problems of
Quantum Gravity} ed. by A. Ashtekar and J. Stachel,
(Birkhauser,Boston,1991).

\end{thebibliography}
\end{document}